\documentclass[prb,twocolumn,superscriptaddress,showpacs,floatfix]{revtex4}

\usepackage{graphicx}% Include figure files
\usepackage{dcolumn}% Align table columns on decimal point
\usepackage{bm}% bold math
\usepackage{color}
\usepackage{gensymb}
\usepackage{multirow}
\newcommand{\nafewo}{NaFe(WO$_4$)$_2$}
\newcommand{\mnwo}{MnWO$_4$}
\newcommand{\tbmno}{TbMnO$_3$}
\newcommand{\DM}{Dzyaloshinskii-Moriya}
\newcommand{\kc}{$\vec{k}_{\text{com}}=(0.5,~0.5,~0.5)$}
\newcommand{\kic}{$\vec{k}_{\text{inc}}=(0.485,~0.5,~0.48)$}
\newcommand{\aaxis}{$\vec{a}$}
\newcommand{\baxis}{$\vec{b}$}
\newcommand{\caxis}{$\vec{c}$}
\newcommand{\fullprof}{\textit{FullProf}}
\newcommand{\easyac}{$\vec{e}_{ac}$}
\newcommand{\qvec}{$\vec{Q}$}
\newcommand{\mubohr}{\,$\mu_{\text{B}}$}
\newcommand{\phmi}{\phantom{-}}
\newcommand{\ptwoc}{\textit{P2/c}}

\begin{document}

\title{Strong magnetoelastic coupling at the transition from harmonic to anharmonic order in NaFe(WO$_4$)$_2$ with 3d$^5$ configuration}

\author{S. Holbein}
\affiliation{II. Physikalisches Institut, Universit\"{a}t zu K\"{o}ln, Z\"{u}lpicher Str. 77, D-50937 K\"{o}ln, Germany}
\affiliation{Institut Laue-Langevin, 71 avenue des Martyrs, F-38042 Grenoble CEDEX 9, France}

\author{M. Ackermann}%
\altaffiliation[Now at  ]{TRUMPF Laser GmbH}\affiliation{Abteilung Kristallographie, Institut f\"ur Geologie und Mineralogie, Universit\"{a}t zu K\"{o}ln, Greinstr. 6, 50939 K\"{o}ln, Germany}

\author{L. Chapon}%
\affiliation{Institut Laue-Langevin, 71 avenue des Martyrs, F-38042 Grenoble CEDEX 9, France}

\author{P. Steffens}%
\affiliation{Institut Laue-Langevin, 71 avenue des Martyrs, F-38042 Grenoble CEDEX 9, France}

\author{A. Gukasov}%
\affiliation{Laboratoire L\'eon Brillouin, CEA/CNRS, F-91191 Gif-sur-Yvette, France}

\author{A. Sazonov}%
\altaffiliation[Now at  ]{Institute of Crystallography, RWTH Aachen University, and J\"ulich Centre for Neutron Science (JCNS) at Heinz Maier-Leibnitz Zentrum, 85747 Garching, Germany}\affiliation{Laboratoire L\'eon Brillouin, CEA/CNRS, F-91191 Gif-sur-Yvette, France}

\author{O. Breunig}%
\affiliation{II. Physikalisches Institut, Universit\"{a}t zu K\"{o}ln, Z\"{u}lpicher Str. 77, D-50937 K\"{o}ln, Germany}

\author{Y. Sanders}%
%\altaffiliation[Now at  ]{???.}
\affiliation{II. Physikalisches Institut, Universit\"{a}t zu K\"{o}ln, Z\"{u}lpicher Str. 77, D-50937 K\"{o}ln, Germany}

\author{P. Becker}%
\affiliation{Abteilung Kristallographie, Institut f\"ur Geologie und Mineralogie, Universit\"{a}t zu K\"{o}ln, Greinstr. 6, 50939 K\"{o}ln, Germany}

\author{L. Bohat\'y}%
\affiliation{Abteilung Kristallographie, Institut f\"ur Geologie und Mineralogie, Universit\"{a}t zu K\"{o}ln, Greinstr. 6, 50939 K\"{o}ln, Germany}

\author{T. Lorenz}%
\affiliation{II. Physikalisches Institut, Universit\"{a}t zu K\"{o}ln, Z\"{u}lpicher Str. 77, D-50937 K\"{o}ln, Germany}

\author{M. Braden}%
\email{braden@ph2.uni-koeln.de}
\affiliation{II. Physikalisches Institut, Universit\"{a}t zu K\"{o}ln, Z\"{u}lpicher Str. 77, D-50937 K\"{o}ln, Germany}

\date{\today}


\begin{abstract}

The crystal structure of the double tungstate NaFe(WO$_4$)$_2$
arises from that of the spin-driven multiferroic MnWO$_4$ by
inserting non-magnetic Na layers. NaFe(WO$_4$)$_2$ exhibits a
three-dimensional incommensurate spin-spiral structure at low
temperature and zero magnetic field, which, however, competes with
commensurate order induced by magnetic field. The incommensurate zero-field
phase corresponds to the condensation of a single irreducible
representation but it does not imply ferroelectric polarization
because spirals with opposite chirality coexist. Sizable
anharmonic modulations emerge in this incommensurate structure,
which are accompanied by large magneto-elastic anomalies, while
the onset of the harmonic order is invisible in the
thermal expansion coefficient. In magnetic fields applied along
the monoclinic axis, we observe a first-order transition to a
commensurate structure that again is accompanied by large
magneto-elastic effects. The large magnetoelastic coupling, a reduction of the
$b$ lattice parameter, is thus associated only with the commensurate order.
Upon releasing the field at low
temperature, the magnetic order transforms to another commensurate
structure that considerably differs from the incommensurate low-temperature phase
emerging upon zero-field cooling. The latter phase, which exhibits a
reduced ordered moment, seems to be metastable.

%\begin{description}
%\item[PACS numbers]
%\end{description}
\end{abstract}

\maketitle

%**********************************************************************************************
%**************************************** NEW SECTION *****************************************
%**********************************************************************************************
\section{\label{sec:intr}INTRODUCTION}
%general

In so-called type-II multiferroics a complex magnetic order directly drives
spontaneous ferroelectric polarization opening the path for possible applications
in data storage or calculation technologies\cite{mf-general}. In most of the
newly discovered multiferroics, in particular in the prototype multiferroic materials $RE$MnO$_3$ with $RE$ for example Tb or Dy \cite{Kimura2003},
the coupled ferroelectric polarization is explained by the inverse \DM\
mechanism~\cite{Katsura2005,Mostovoy,Sergienko2006}. While in typical magnetic systems
antisymmetric coupling arises from a low crystal symmetry and induces spin canting,
an intrinsically non collinear magnetic structure can drive a structural distortion
and thereby enhance or even create antisymmetric coupling. However, only if this structural distortion also develops
a macroscopic ferroelectric polarization, the system is multiferroic. The antisymmetric coupling
is induced by spin-orbit-coupling and therefore much smaller than the dominant symmetric exchange
interaction. In consequence, the ferroelectric polarization values induced by the inverse \DM\ mechanism
are typically small\cite{mf-general}, two or more orders of magnitude smaller than in a normal ferroelectric.
Modifying the symmetric and isotropic exchange seems more promising to obtain multiferroics with
large ferroelectric polarization, and it was proposed \cite{sergienko} and experimentally confirmed\cite{ishiwata} that orthorhombic $RE$MnO$_3$ with
smaller $RE$ exhibit such a large multiferroic polarization basing on exchange striction. For smaller
$RE$ the magnetic structure changes from the incommensurate cycloid observed for Tb or Dy to a commensurate
up-up-down-down structure (called E-type), in which the scalar product of neighboring moments entering the symmetric exchange
even changes sign. This exchange-striction based magnetoelectric coupling \cite{sergienko} not only explains the static coupling in the multiferroic
phase of the E-type ordered $RE$MnO$_3$, but it also constitutes the dominant dynamic magnetoelectric coupling resulting in
the strongest electromagnon modes\cite{elm1,elm2} in the $RE$MnO$_3$ with larger $RE$ that exhibit the incommensurate cycloidal order.
The distinct multiferroic phases in
$RE$MnO$_3$ thus arise from the competition between incommensurate cycloid and commensurate up-up-down-down orders, and this competition
is controlled through the structural distortions following the $RE$ ionic radius. Here,
we investigate \nafewo , which also exhibits a competition between incommensurate cycloid and up-up-down-down
phases, and which, therefore, may help understanding the complex magnetoelastic coupling in such phase diagrams. Also in \nafewo\ we find
rather strong magnetoelastic coupling, however without any ferroelectric polarization so that none of the
phases of \nafewo\ is multiferroic.

The discovery of a spin-driven multiferroic phase in \mnwo\ in
2006~\cite{Heyer2006,Arkenbout2006,Taniguchi2006} motivated the
research for multiferroicity in other materials of the tungstate
family~\cite{diss_jodlauk}. The magnetic moments in \mnwo\ develop
a spin spiral at low temperature which is the driving force of
the ferroelectric polarization, explained by the inverse \DM\
mechanism~\cite{Kimura2003,Katsura2005,Mostovoy,Sergienko2006}. The metal
ion $M^{2+}$ in \textit{M}WO$_4$ can be substituted by a magnetic
ion with the same valency or by a combination of mono- and
trivalent ions. The resulting compounds often develop a simple
collinear antiferromagnetic structure, which is the case for
FeWO$_4$~\cite{Cid1968}, CoWO$_4$, NiWO$_4$,
CuWO$_4$~\cite{Weitzel1976} and NaCr(WO$_4$)$_2$~\cite{Nyam-ochir2008}. No
electric polarization was observed in these
compounds~\cite{diss_jodlauk}.

\begin{figure}[!t]
    \centering
    \begin{minipage}[b]{0.45\textwidth}
        \centering
        \includegraphics[width=0.95\textwidth]{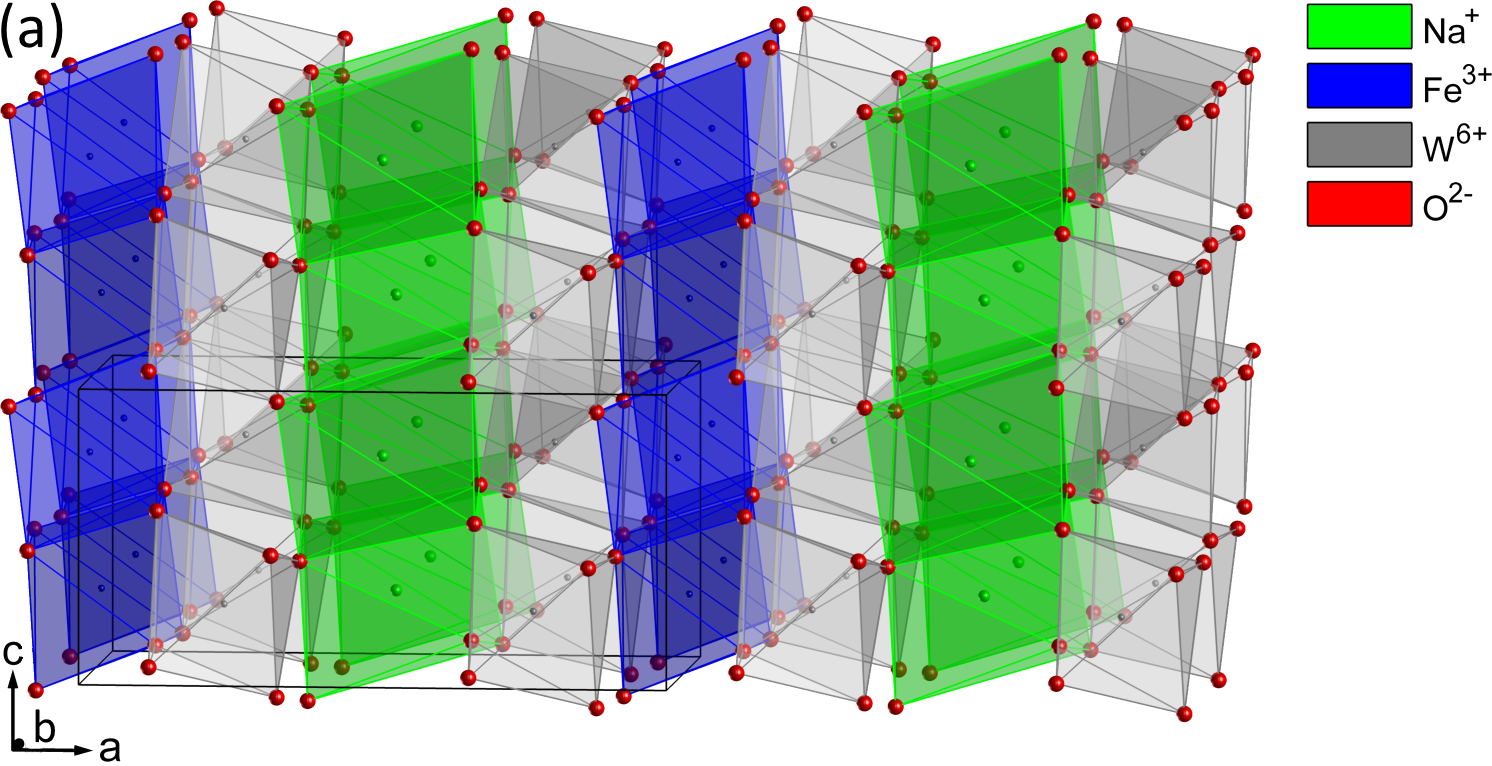}
    \end{minipage}
    \begin{minipage}[b]{0.25\textwidth}
        \centering
        \includegraphics[width=0.95\textwidth]{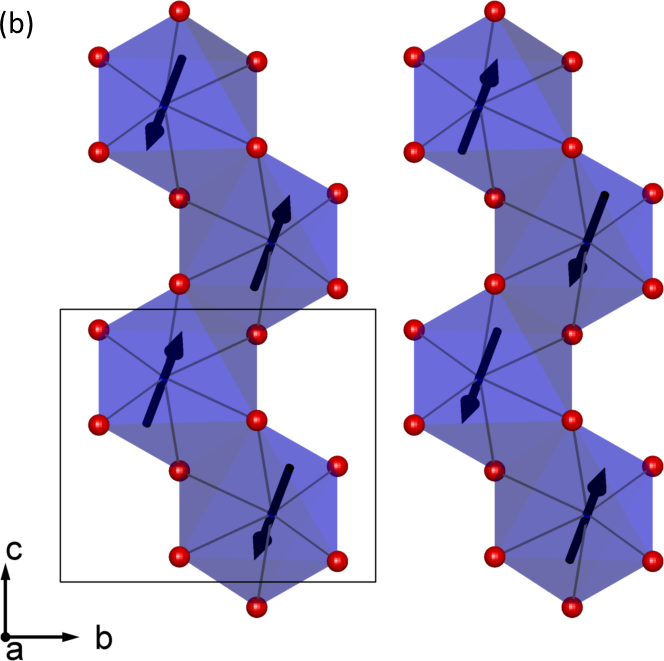}
    \end{minipage}
            \caption{(a) Crystal structure of \nafewo. Tungsten, iron and sodium ions are surrounded by oxygen octahedra. These octahedra form edge-sharing chains along \caxis\ and order in planes perpendicular to \aaxis. (b) Zig-zag chains of iron octahedra propagate along \caxis. The arrows show the magnetic up-up-down-down structure with propagation vector \kc~\cite{Nyam-ochir2008}. The black box indicates the crystallographic unit cell.}
            \label{fig:nafewo_nuc}
\end{figure}

%crystal and magnetic structure
The crystal structure of the double tungstate \nafewo\ can be
described in the monoclinic space group \textit{P2/c} with lattice
parameters $a=9.88$\,\AA, $b=5.72$\,\AA, $c=4.94$\,\AA\ and a
monoclinic angle of $\beta=90.33$\degree~\cite{Klevtsov1970}. Na$^+$
and Fe$^{3+}$ ions are surrounded by edge-sharing [O$_6$] octahedra. These
octahedra form zig-zag chains along \caxis\ and align in planes
parallel to the $bc$ plane. The crystal structure is shown in
Figure~\ref{fig:nafewo_nuc}(a). Layers containing [NaO$_6$]
and [FeO$_6$] octahedra, respectively, are separated by layers that
contain [WO$_6$] octahedra only. Due to the insertion of the Na
planes the unit cell of \nafewo\ is doubled along \aaxis, with
respect to the natural wolframites \mnwo\ and FeWO$_4$, which
otherwise crystallize in the same space
group~\cite{Weitzel1976,Cid1968}. Therefore, the magnetic
interaction between Fe$^{3+}$ is considerably weakened along
\aaxis \ resulting in a lower N\'eel temperature.

Similar to the case of \mnwo, the magnetic Fe$^{3+}$ ions in \nafewo\ form
zig-zag chains along the $c$ axis (see
Fig.~\ref{fig:nafewo_nuc}(b)). In spite of the long distance
between the Fe$^{3+}$ ions in adjacent layers, \nafewo\ develops a
three-dimensional magnetic structure at temperatures below 4\,K.
The analysis of neutron powder diffraction yielded a collinear
antiferromagnetic structure with magnetic moments aligned parallel
to the $a$ axis~\cite{Nyam-ochir2008}. The magnetic reflections
were indexed with a commensurate propagation vector \kc \ that corresponds to an up-up-down-down magnetic arrangement along the chains, which
can be explained by a dominating next-nearest neighbor magnetic interaction within the chains.

% Nyam-Ochir \textit{et al.} claim that
%the antiferromagnetic superexchange coupling along \baxis\
%and \caxis\ causes the doubling of the magnetic unit cell along
%these directions, whereas the mechanism for the coupling along the
%extended $a$ axis remains unclear~\cite{Nyam-ochir2008}.

In this article, we present a comprehensive investigation of the
magnetic properties of \nafewo\ in zero field and in magnetic
fields applied along the monoclinic axis \baxis\ by combining
various macroscopic and neutron diffraction techniques on single
crystals. We show that the zero-field magnetic structure is more
complex than previously proposed\cite{Nyam-ochir2008}, because it
develops an incommensurate spin spiral, which, however, does not result in a multiferroic phase. Most interestingly, there
are several phase transitions associated with the emergence of
anharmonic components, whose signatures in some macroscopic
properties (thermal expansion) are even larger than those associated with the
onset of magnetic order in zero magnetic field.

%**********************************************************************************************
%**************************************** NEW SECTION *****************************************
%**********************************************************************************************

\section{\label{sec:sym}Symmetry analysis}

The magnetic symmetry of the system has been derived by applying
representation analysis~\cite{Bertaut1968}. The
crystallographic structure of \nafewo\ can be described in the
space group \ptwoc\ (No.~13). The magnetic Fe$^{3+}$ ions are
located at the special Wyckoff site $2e$ at $(0,~0.670,~1/4)$,
which has two-fold symmetry.
%commensurate propagation vector

Nyam-Ochir \textit{et al.} were able to describe the magnetic
neutron powder data with a commensurate propagation vector of
\kc~\cite{Nyam-ochir2008}. The
corresponding little group $G_{\vec{k}_{\text{c}}}$ is identical to the space group \ptwoc. It
contains one two-dimensional irreducible representation
$\Gamma_1$. In the case of the commensurate propagation vector,
$\vec{k}_{\text{c}}$ and $-\vec{k}_{\text{c}}$ are equivalent and the star of
$\vec{k}_{\text{c}}$ consists of one vector. The character table and the
corresponding symmetry conditions for the magnetic moments are
given in Table~\ref{tab:nafewoCharC}. The two-dimensional
representation allows the two symmetry-connected moments in the
crystallographic unit cell to be either collinear or canted. For a
given moment $(u,~v,~w)$, the second moment in the unit cell can
align according to the four possibilities: $(u,v,w)$,
$(u,\bar{v},w)$, $(\bar{u},v,\bar{w})$ and
$(\bar{u},\bar{v},\bar{w})$. The low-temperature commensurate
magnetic structure AF1 in \mnwo\ is also described by this little
group~\cite{Lautenschlager1993}.

\begin{table}
        \centering
        \renewcommand{\arraystretch}{1.3}
\caption{Character table and symmetry conditions of the little Group $G_{\vec{k}_{\text{c}}}=$\ptwoc, \kc.}
\begin{tabular}[b]{l||c|c|c|c|l|l}
     & 1& 2 & $\bar{1}$ & c & $(x,y,z)$ & $(\bar{x},\bar{y},\bar{z})$ \\
    \hline
    \multirow{2}{*}{$\Gamma_1$} & \phmi1 \phmi0 & \phmi1 \phmi0 & \phmi0 \phmi1 & \phmi0 \phmi1 & \multirow{2}{*}{ $(u,v,w)$} & \multirow{2}{*}{$(p,q,r)$}\\
                                         & \phmi0 \phmi1 & \phmi0 -1     & \phmi1 \phmi0 & -1 \phmi 0 & & \\
\end{tabular}
\label{tab:nafewoCharC}
\end{table}
%incommensurate propagation vector

Neutron diffraction studies on a single crystal of \nafewo\ reveal an
incommensurate magnetic propagation vector of the form
$\vec{k}_{\text{ic}}=(\delta_H,0.5,\delta_L)=$ $(0.485,~0.5,~0.48)$ (see
Section~\ref{sec:mic-prop}). Note, however, that this incommensurate vector is very close to $(0.5,~0.5,~0.5)$, so that the magnetic structure still locally resembles the up-up-down-down sequence shown in Fig.~\ref{fig:nafewo_nuc}(b).

With the incommensurability along
\aaxis* and \caxis* the little group of the magnetic structure
changes to $G_{\vec{k}_{\text{ic}}}=\{1,c\}$. It contains two
one-dimensional irreducible representations $\Gamma_{1,2}$. The
character table and the corresponding symmetry conditions for the
magnetic moments are given in Table~\ref{tab:nafewoCharIC}. In the
case of the incommensurate propagation vector, $\vec{k}_{\text{ic}}$ and
$-\vec{k}_{\text{ic}}$ are not equivalent and the star of $\vec{k}_{\text{ic}}$
contains two vectors. Because the $c$ glide plane connects the two
Fe sites in the unit cell and since $c$ belongs to
$G_{\vec{k}_{\text{ic}}}$, both sites thus belong to one orbit and can be
described by three complex parameters $u,v,w$, cf.
Table~\ref{tab:nafewoCharIC}. The incommensurate magnetic
structures AF2 and AF3 of \mnwo\ are also described in this
little group~\cite{Lautenschlager1993}.

The two irreducible representations $\Gamma_1$ and $\Gamma_2$ are
thus described by three complex amplitudes $u, v, w$,  whose six
independent parameters can be reduced to five by arbitrarily
choosing one phase, e.g. $u=|u|$. Further insight can be gained by
magnetic superspace symmetry analysis, which takes into account
additional symmetry elements not keeping $\vec{k}_{\text{ic}}$ invariant
\cite{perez}. The superspace analysis yields further constraints
to the five remaining parameters of each symmetry by fixing the
phases, see Table~\ref{tab:nafewoCharIC} \cite{Urcelay-Olabarria2013}. Because
$u$ and $w$ always have the same phase that differs from that of $v$
by $\pm\frac{\pi}{2}$, a spiral magnetic structure emerges at each of the two Fe-sites, but these two spirals have opposite chirality, which will be essential for the understanding of the absence of a multiferroic phase, see below.

\begin{table}
        \centering
        \renewcommand{\arraystretch}{1.3}
\caption{Character table and symmetry conditions  of the little
Group $G_{\vec{k}_{\text{ic}}}=Pc$,
$\vec{k}_{\text{ic}}=(\delta_H,~0.5,~\delta_L)$ with $a=e^{-i2\pi \cdot
\delta_L \cdot r_z}=e^{-i2\pi \cdot 0.24}$.}
\begin{tabular}[b]{l||l|l|l|l|l|l}
    & 1& c & $(x,y,z)$ & $(x,\bar{y},z+1/2)$ & super-space symmetry \\
    \hline
    $\Gamma_1$              & \phmi1 & -a  & $(u,v,w)$ & $a\cdot(u,\bar{v},w)$ & $u,w$ imaginary, $v$ real\\
    $\Gamma_2$              & \phmi1 & \phmi a       & $(u,v,w)$ & $a\cdot(\bar{u},v,\bar{w})$ & $u,w$ real, $v$ imaginary\\
\end{tabular}
\label{tab:nafewoCharIC}
\end{table}

%**********************************************************************************************
%**************************************** NEW SECTION *****************************************
%**********************************************************************************************
\section{\label{sec:exp}Experimental Methods}
%crystals
The macroscopic and microscopic properties presented in this article were measured on single crystals of \nafewo. The crystals were grown from sodium poly-tungstate flux (starting ratio Na$_2$W$_2$O$_7$~:~\nafewo~=~3~:~2, with excess of WO$_3$) by the top seeded solution growth technique within the temperature range from 1172 to 1163\,K. During typical growth periods of four weeks, dark green single crystals of up to 1\,cm$^3$ volume and well-developed morphology were obtained. The neutron scattering experiments have been performed on two samples of sizes $13\times8\times2$\,mm$^3$ and $6\times7\times2$\,mm$^3$, respectively. Macroscopic measurements were performed on smaller pieces of the same batch.

%macroscopic
The magnetization was measured using a commercial superconducting quantum interference device (SQUID) magnetometer as a function of temperature from 1.8 to 300~K in magnetic fields up to 7~T applied along the principal crystallographic directions of the crystals. The specific heat ($c_p$) was measured by the thermal relaxation-time method using a home-built calorimeter. The temperature and magnetic-field dependent length changes $\Delta L_b(T,B)$ were measured with a home-built capacitance dilatometer along the $b$ axis. By numerically derivating the relative length changes with respect to temperature or magnetic field, the thermal expansion ($\alpha=1/L_b^0 \;\partial \Delta L_b/\partial T$) or magnetostriction ($\lambda=1/L_b^0 \;\partial \Delta L_b/\partial B$) coefficients are obtained. The calorimeter (dilatometer) was attached to the $^3$He pot in the high-vacuum chamber of a $^3$He-cryostat and $c_p$ ($\Delta L_b$) was measured in the temperature range from about 300~mK to 10~K in magnetic fields up to 17~T applied along the monoclinic $b$ axis of the single-crystalline samples.

%microscopic
Neutron diffraction experiments have been performed at different instruments. The crystal and magnetic structure was investigated at the four-circle diffractometer D10 (ILL, Grenoble) at 12\,K and 1.75\,K, respectively. The $Q$ and temperature dependence of the magnetic propagation vector was studied at the triple-axis spectrometers IN3 and IN14 (both ILL, Grenoble) using different crystal orientations. Finally, the high-field phases in magnetic fields applied along \baxis\ were investigated at the four-circle diffractometer 6T2 with lifting-counter and vertical cryomagnet (LLB, Saclay).

%**********************************************************************************************
%**************************************** NEW SECTION *****************************************
%**********************************************************************************************
\section{\label{sec:mac}Macroscopic measurements}
\subsection{\label{sec:mac-squit}Magnetization}
\begin{figure}[!t]
    \centering
    \includegraphics[width=0.45\textwidth]{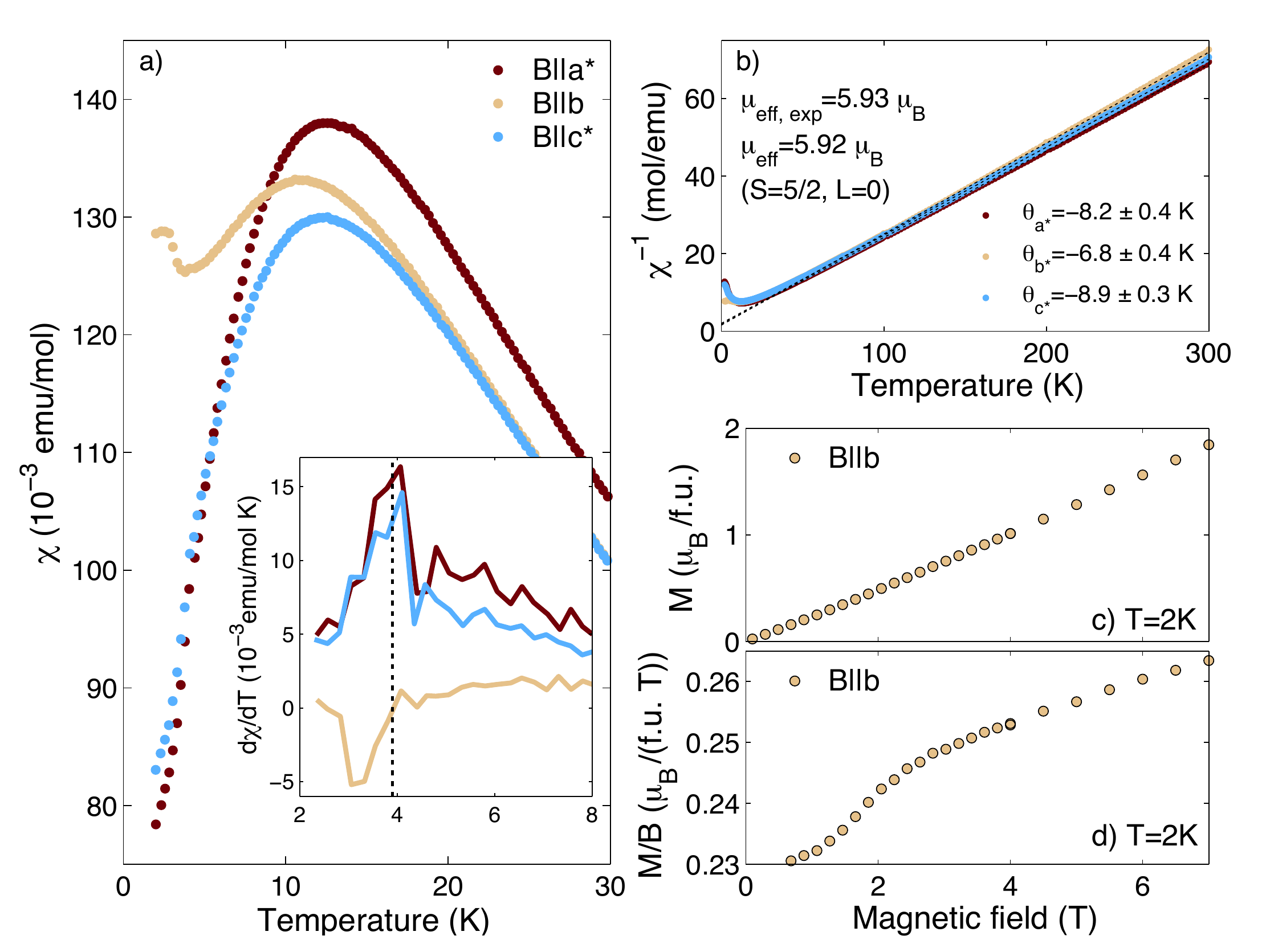}
    \caption{Magnetic susceptibility of \nafewo\ along principal crystallographic directions: (a) low temperature behaviour and (b) Curie Weiss fit to the high temperature part. The inset in (a) shows the derivative of the susceptibility with respect to the temperature. A magnetic field of 0.1\,T was applied to the sample. (c) Magnetization on \nafewo\ for magnetic fields applied along \baxis\ on decreasing fields. (d) Magnetization per applied magnetic field along the same direction.}
    \label{fig:nafewo_SQUID}
\end{figure}
Figures~\ref{fig:nafewo_SQUID}(a-c) show the magnetic
susceptibility of \nafewo\ along the principal crystallographic
directions measured on a single crystal. The high-temperature part
was fitted by the inverse Curie-Weiss function
$\chi_{\text{m}}^{-1}=C^{-1}(T+\theta)$ yielding Weiss temperatures $\theta$ ranging from about -6.8 to -8.9~K
depending on the field direction, see Fig.\ref{fig:nafewo_SQUID}(b), and an average effective magnetic moment
$\mu_{\text{eff,exp}}=5.93$\mubohr , in very good agreement with
%$\theta=-8.2(1.1)$\,K for measurement directions within the monoclinic plane and an effective magnetic
% moment of $\mu_{\text{eff,exp}}=5.932(15)$\mubohr, which matches perfectly
the expected value of
$\mu_{\text{eff}}=2\sqrt{S(S+1)}$\mubohr\ $=5.92$\mubohr\ for a spin-only moment of
Fe$^{3+}$ with $S=5/2$. The results agree with values determined
from powder samples~\cite{Nyam-ochir2008}.

An expanded view of the low-temperature range of the susceptibility is shown in Fig.~\ref{fig:nafewo_SQUID}(a). All three $\chi(T)$ curves show broad
maxima around 14\,K, which signal the occurrence of strong magnetic correlations in the temperature range well above
the transition temperature $T_{\rm N}\simeq 3.9$~K where long-range magnetic order sets in. This value of $T_{\rm N}$ is obtained from the extrema of the temperature derivatives of $\chi(T)$ and is indicated by a dashed line in the inset of Fig.~\ref{fig:nafewo_SQUID}(a). The occurrence of strong magnetic correlations above $T_{\rm N}$ is typical for low-dimensional magnetic systems and is naturally expected for \nafewo , which consists of two-dimensional layers of magnetic Fe$^{3+}$ spin chains that are weakly coupled along the perpendicular direction \aaxis.

Below $T_{\rm N}$, the susceptibilities for fields along \aaxis* and \caxis* further decrease, whereas $\chi_b(T)$ even slightly increases.
This anisotropic behavior already starts near the maxima of $\chi_i(T)$, i.e. well above $T_{\rm N}$, and it indicates that the magnetic moments are preferentially aligned perpendicular to the $b$ axis. As will be seen below, within the $ac$ planes, the magnetic moments align approximately along the axis bisecting \aaxis* and \caxis*, which explains the very similar temperature dependencies of $\chi_i$ for magnetic fields applied along these two directions.

Finally, in Figure~\ref{fig:nafewo_SQUID}(c) we show the induced magnetization
for fields up to $B=7$\,T $\parallel$ \baxis. The magnetization almost linearly increases with field and reaches about 2\,$\mu_{\rm B}$/f.u., i.e.\ about $40\%$ of the expected saturation magnetization of the $S=5/2$ spin moments of Fe$^{3+}$.
However, a closer inspection of the magnetization per field (cf. Figure~\ref{fig:nafewo_SQUID}(d)) reveals an anomaly at
about 2\,T indicating a magnetic reorientation, which will be discussed in detail below.

\subsection{\label{sec:mac-heat}Specific heat}

The specific heat of \nafewo\ measured at low temperatures is displayed in Figure~\ref{fig:02-mac}(a) for different magnetic fields
applied along \baxis. In general, the values were
determined during a heating run by step-wise heating the sample (red data points). In addition, we measured $c_p$ for various fields also during a cooling run by successively decreasing the base temperature (black data points), but in none of these measurements a clear temperature hysteresis could be resolved.

In zero field and in 1\,T, the specific heat shows a rather broad maximum in $c_p/T=\partial S/\partial T$ at 4\,K signalling an inflection point of the temperature dependence of the (magnetic) entropy. This feature corresponds nicely to the magnetic transition temperature $T_{\rm N}\simeq 3.9$~K as
determined from the magnetization. However, the weakness of this feature also reveals that instead of a sharp transition the 3-dimensional correlations develop rather gradually in NaFe(WO$_4$)$_2$. Again, this behavior can be naturally explained by the weakly coupled 2-dimensional magnetic planes in \nafewo . Upon lowering the temperature, the in-plane magnetic correlations continuously evolve such that the magnetic entropy continuously freezes already well above $T_{\rm N}$, and the 3-dimensional ordering only causes a weak additional decrease of magnetic entropy.

Above 2\,T, the feature at the transition temperature sharpens significantly and its shape indicates a first-order phase transition, but as already mentioned there is essentially no temperature hysteresis detectable. Moreover, the total entropy change in the temperature range from 0.3 to 10\,K (see Fig.~\ref{fig:02-mac}(b)) only amounts to about 60\% of the expected full magnetic entropy $S_{\text{mag}}= N_{\rm A}k_{\rm B} \ln(2S+1)\simeq 14.9$\,J/mol/K of an $S=5/2$ system. In fact, this total entropy change hardly varies from zero field up to 6~T, although the entropy decrease at the transition sharpens up above 2\,T. For all fields studied, this entropy change remains below 20\% of the expected total magnetic entropy, which once again emphasizes the importance of short-range correlations persisting well above $T_{\rm N}$. Above 8\,T,  the total entropy change as well the transition temperature systemically decrease with further increasing field and the antiferromagnetic order is fully suppressed above about 15\,T.

\subsection{\label{sec:mac-te}Thermal expansion and magnetostriction}
%thermal expansion

\begin{figure}[t]
    \begin{minipage}[b]{0.19\textwidth}
        \includegraphics[width=\textwidth]{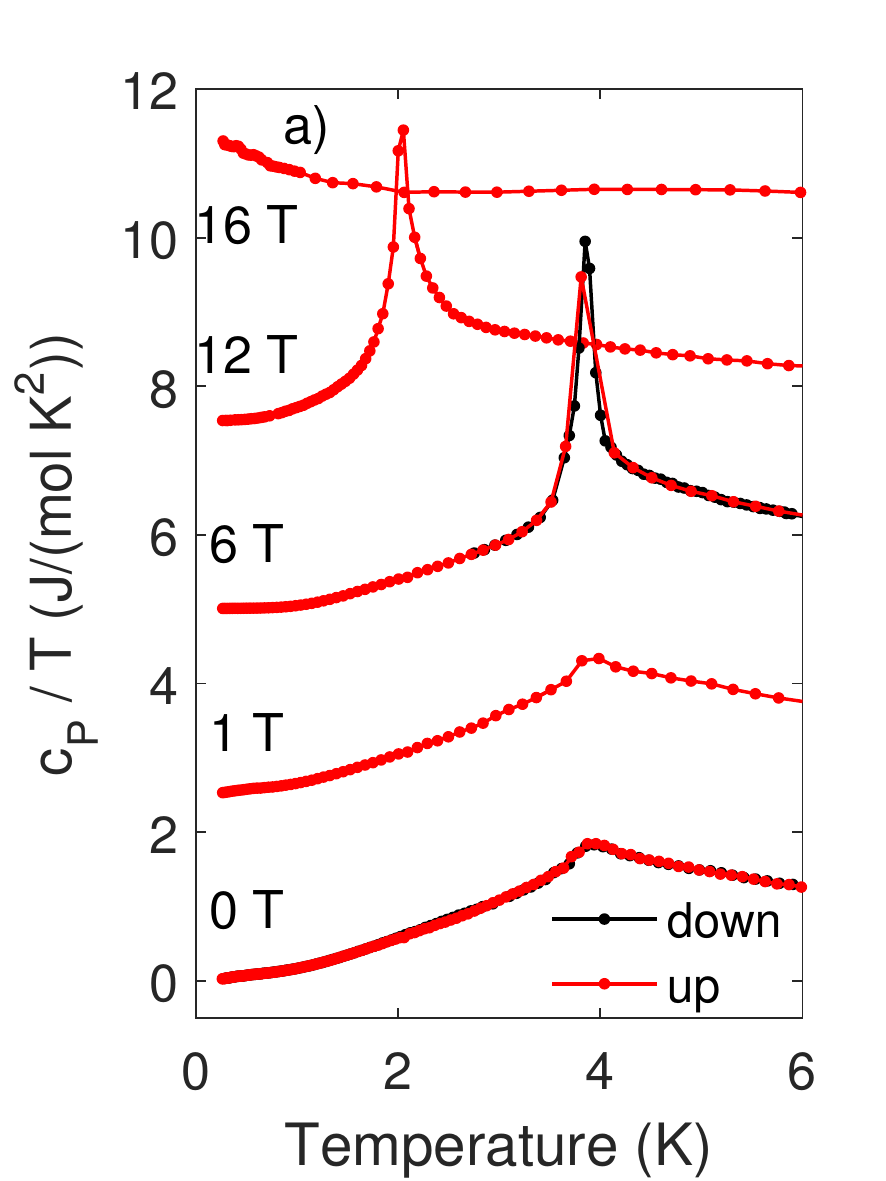}
    \end{minipage}
    %\hfill
    \begin{minipage}[b]{0.19\textwidth}
        \includegraphics[width=\textwidth]{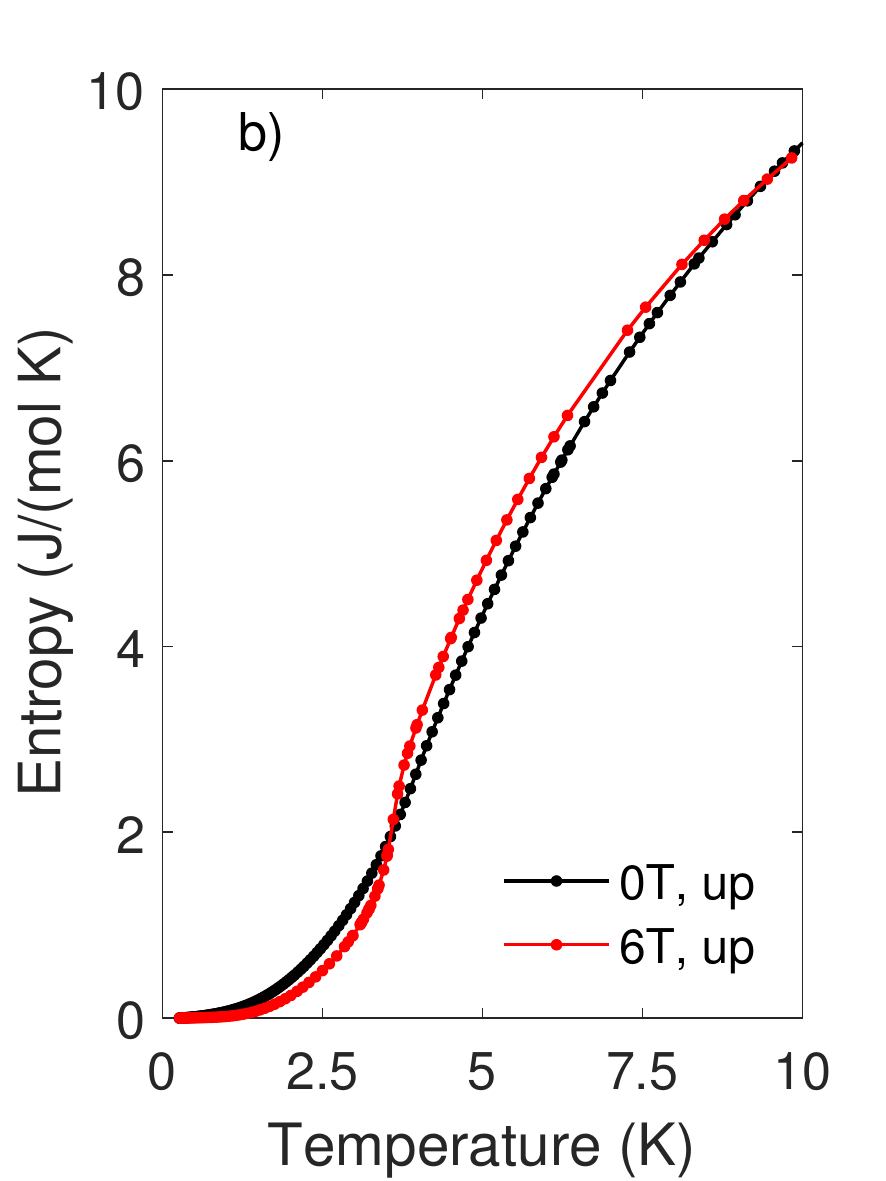}
    \end{minipage}
    \begin{minipage}[b]{0.19\textwidth}
        \includegraphics[width=\textwidth]{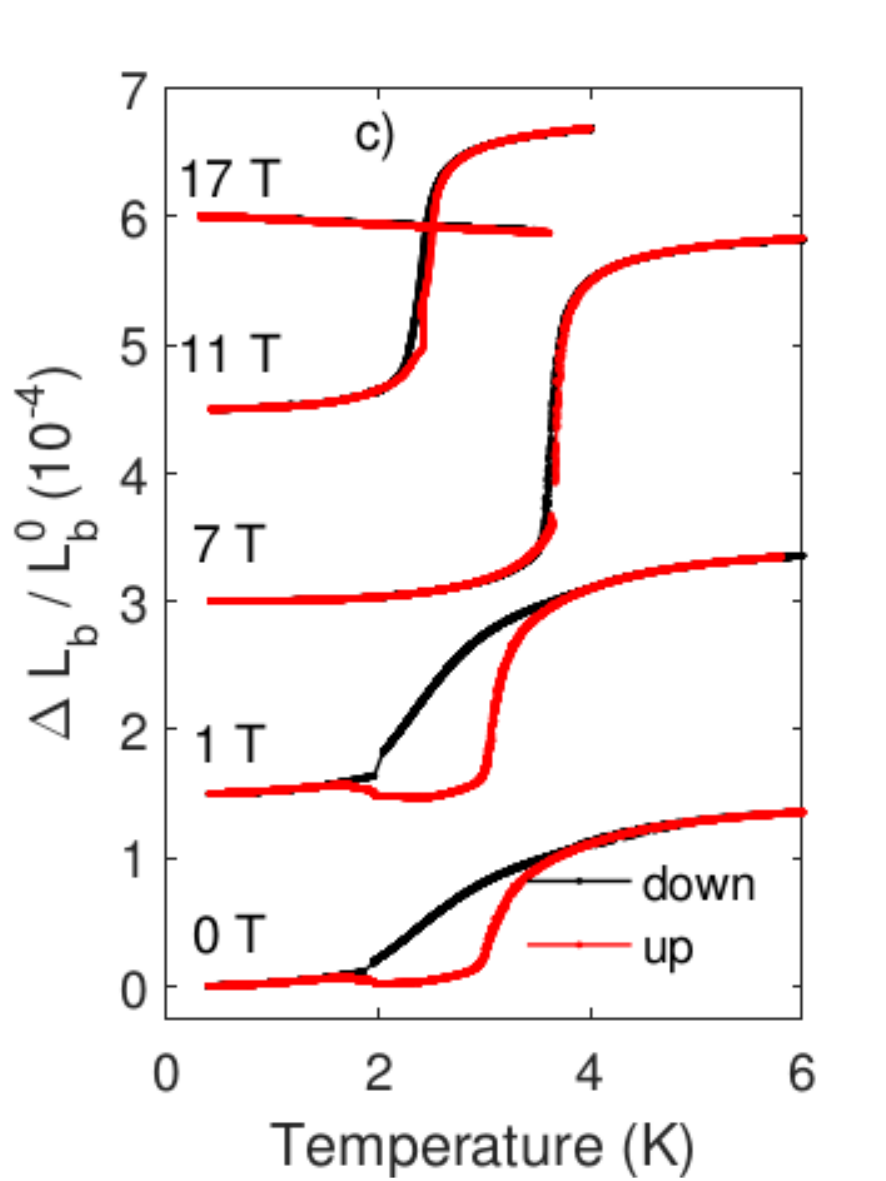}
    \end{minipage}
    \begin{minipage}[b]{0.19\textwidth}
        \includegraphics[width=\textwidth]{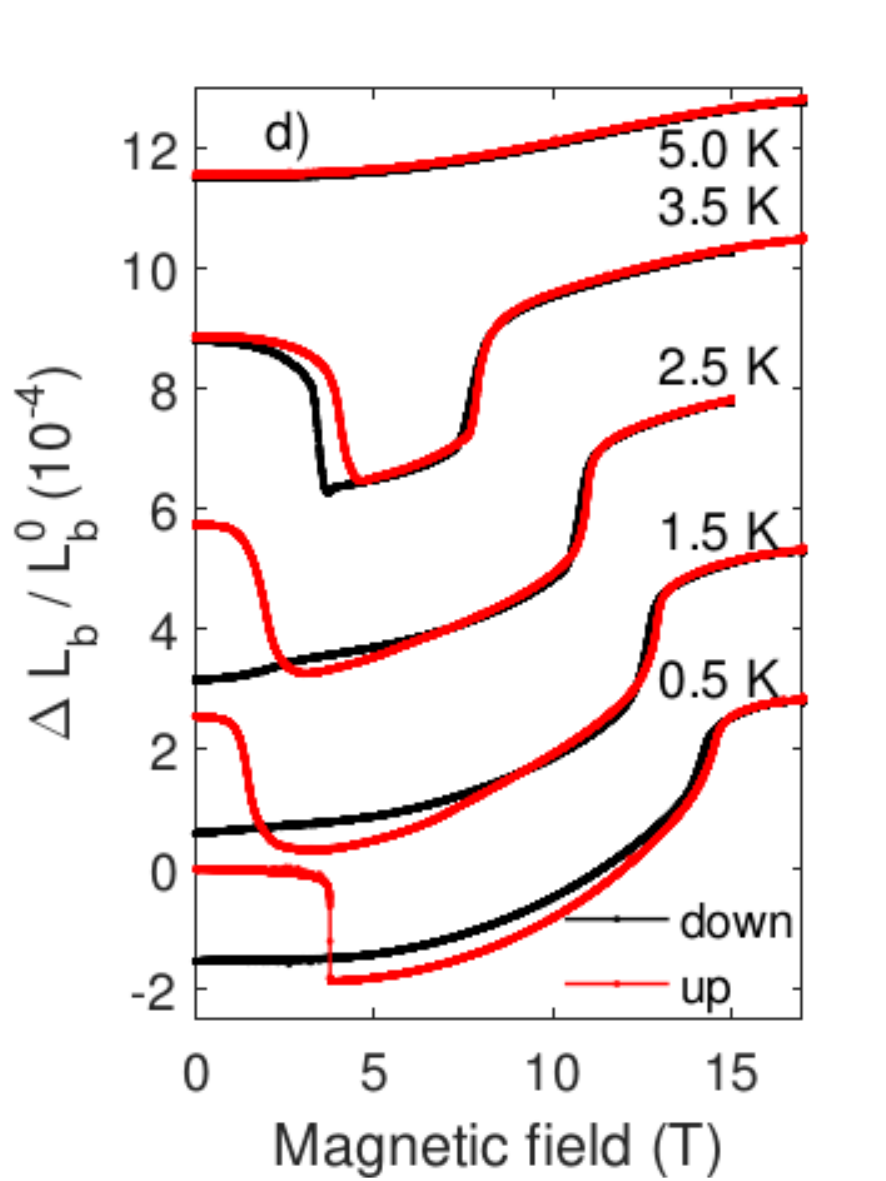}
    \end{minipage}
    %\hfill
    %\begin{minipage}[b]{0.25\textwidth}
    \caption{Specific heat (a), entropy (b), thermal expansion (c) and magnetostriction (d) of \nafewo\ for magnetic fields applied along \baxis. The length changes in (c) and (d) were measured along the monoclinic axis \baxis. The specific heat and thermal expansion data were obtained upon decreasing and increasing temperature (black and red curves, respectively). For the magnetostriction, the sample was cooled in zero-field before the field was increased and decreased at constant temperature (red and black curves, respectively). The curves are separated by a constant offset: (a) $2.5\cdot10^{-4}$\,J/mol/K$^{2}$, (c) $1.5\cdot10^{-4}$ and (d) $2.5\cdot10^{-4}$.}
    \label{fig:02-mac}
    %\end{minipage}
\end{figure}

Figures~\ref{fig:02-mac}(c) and (d) show the thermal expansion and magnetostriction of \nafewo\ along the monoclinic axis \baxis\ for magnetic fields applied along the same direction. Because of the very strong magnetoelastic effects of the order of $10^{-4}$ the relative length changes $\Delta L_b(T,B)/ L_b^0 = \frac{\Delta b}{b}$ are displayed here instead of the corresponding temperature or field derivatives $\alpha$ or $\lambda$. All curves were measured upon continuously increasing (red lines) and decreasing (black lines) either the temperature at constant $B$ or the field at constant $T$. In magnetic fields above 2\,T, the thermal expansion measurements confirm some of the basic observations from the specific heat measurements. The phase transitions cause very large and sharp changes of $\Delta L_b(T)/L_b^0$, whose shape and magnitude are typical for first-order phase transitions, but no systematic temperature hysteresis is present in this field range.

For zero field and in 1~T, however, the thermal expansion data reveal a systematically different behavior compared to the $c_p(T)$ measurements. Upon cooling, $\Delta L_b(T)$ continuously decreases down to about 2~K, but then the slope abruptly changes to a moderate decrease of $\Delta L_b(T)$ upon further cooling to the minimum temperature of 300~mK. In the heating run,
$\Delta L_b(T)$ reversibly follows the cooling curve only up to about 1.8~K. Then, on further heating, $\Delta L_b(T)$ first shows a broad minimum around 2.5~K, which is followed by an abrupt steep increase at 3~K and around 3.5~K the heating curve of $\Delta L_b(T)$ finally meets the previous cooling curve and no further hysteresis is observed. Most surprisingly, the hysteresis and the anomalies of the low-field thermal expansion are observed in a temperature range where the corresponding $c_p(T)$ curves are fully reversible without any anomalies. In contrast, in the thermal expansion data no anomaly shows up at the N\'eel temperature $T_{\rm N}\simeq 4$~K. As will be discussed below, this is related to the fact that an incommensurate, anharmonic low-field phase develops in \nafewo\ and $\Delta L_b(T)$ scales with both the variation of incommensurability and with the anharmonicity of magnetic order.

The magnetostriction $\Delta L_b(B)$ measured after zero-field cooling is displayed for selected temperatures in Fig.~\ref{fig:02-mac}(c). At 0.5~K, a large discontinuous contraction of $\Delta L_b(B)$ takes place at $B_{c1}^{up}\simeq 3.8\;$T, which is followed by a continuous expansion up to about $ B_{c2}^{up}\simeq 14.5\;$T where an abrupt expansion occurs and above $ 15\;$T a saturation of $\Delta L_b(B)$ sets in. With decreasing field, the upper transition is shifted by $\simeq 0.3\;$T towards lower fields and reverses the abrupt length change, whereas the lower transition is absent. On increasing temperature, the abrupt length change at the upper transition systematically increases, whereas the transition field and the hysteresis width decrease to $B_{c2}^{up}\simeq 9.7\;$T and $B_{c2}^{up}-B_{c2}^{down}\simeq 0.1\;$T, respectively, at $T=3\;$K. The magnetostriction anomalies at the upper transition well agree with the corresponding anomalies of the thermal-expansion and the specific-heat data in the $(B,T)$ plane and reveal that the magnetic-ordering transition of \nafewo\ in finite magnetic fields is a first-order transition and that this antiferromagnetic order is fully suppressed above about $ 15\;$T. The absence/presence of the lower transition reveals that there are metastable phases in the low-field low-temperature range. The field range of these metastable phases shrinks with increasing temperature and seems to vanish around 3~K, where the magnetostriction $\Delta L_b(B)$ indicates a first-order low-field transition with a pronounced hysteresis  $B_{c1}^{up}-B_{c1}^{down}\simeq 1\;$T.
Note, however, that $\Delta L_b(B)$ is not fully reversible at this transition and, moreover, additional hysteresis effects are also present in $\Delta L_b(B,T)$ over a wider field and temperature range. These effects most probably arise from coexisting phases due to incomplete first-order transitions.

\subsection{\label{sec:mac-phase}Phase diagram}
\begin{figure}[!t]
    \centering
    \includegraphics[width=0.37\textwidth]{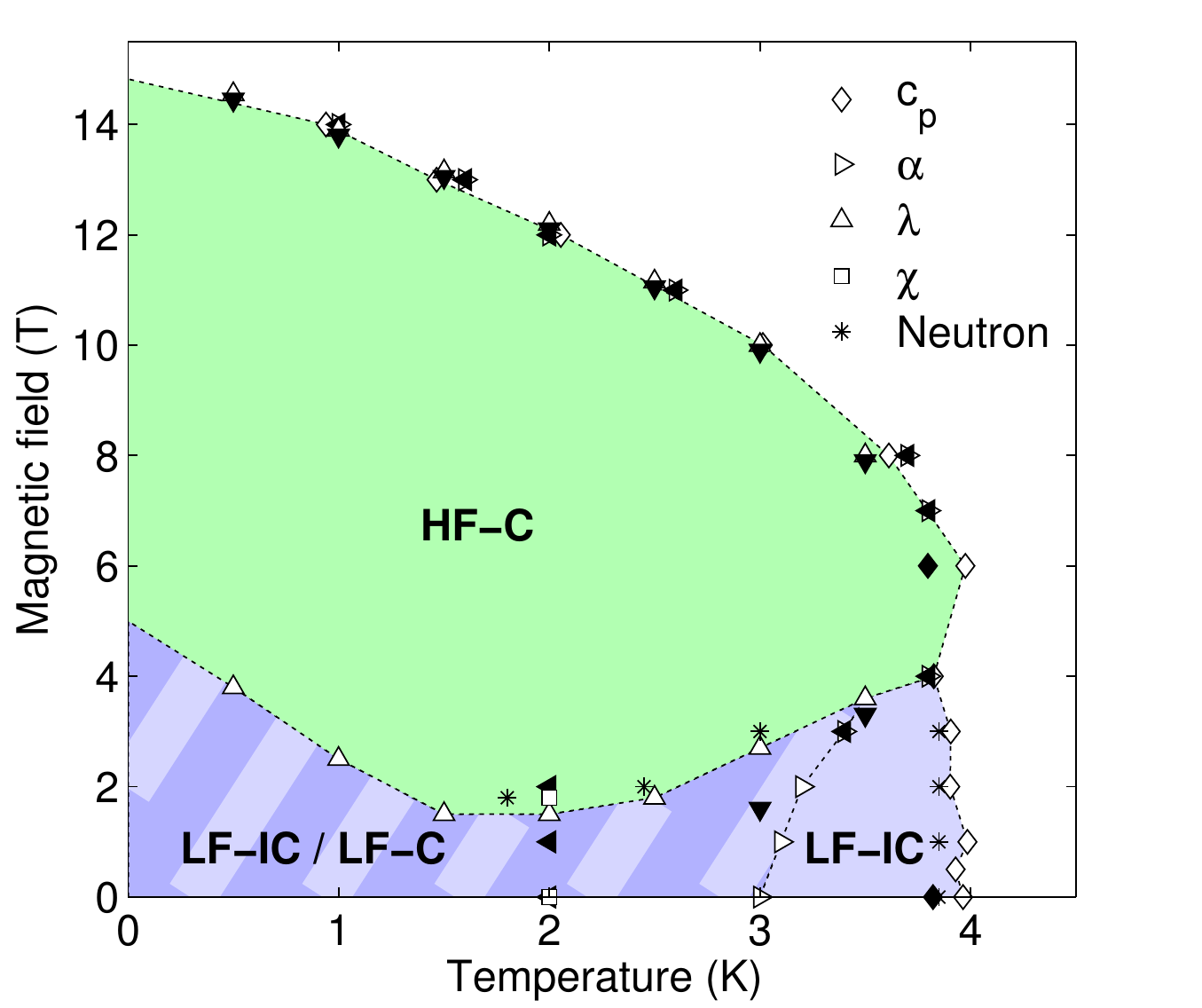}
    \caption{Magnetic phase diagram of \nafewo\ for applied magnetic field along \baxis. The transition temperatures derived from measurements of the specific heat ($c_p$), thermal expansion ($\alpha$), magnetostriction ($\lambda$), magnetization ($\chi$) and neutron diffraction are indicated upon heating (open symbols) and cooling (filled symbols). We distinguish three different phases: low-field incommensurate (LF-IC), low-field commensurate (LF-C) and high-field commensurate (HF-C).}
    \label{fig:03-phase}
\end{figure}

The basic features of the magnetic phase diagram of \nafewo\ are summarized in Fig.~\ref{fig:03-phase}. Open and filled symbols denote transition fields or temperatures that were obtained from the anomalies of the various macroscopic quantities ($\chi$, $c_p$, $\alpha$, $\lambda$), either upon increasing or decreasing the magnetic field or temperature, respectively. In addition, important microscopic information is included that is obtained from the neutron diffraction measurements, which will be discussed below. The ordered phases of \nafewo\ cover a field and temperature range below about 4~K and 15~T, which can be divided into three different regions. Above about 4~T, there is a high-field ordered phase HF-C with commensurate magnetic order, whereas the low-field region further splits into two regions. Below about 3~K, there are two low-field phases LF-IC and LF-C with incommensurate and commensurate magnetic order, respectively. The LF-IC phase is reached upon zero-field cooling, while the LF-C phase is observed after successively ramping the magnetic field up and down at low temperature. In the temperature range between 3 and 4~K, another incommensurate low-field phase LF-IC* is found, that differs from the LF-IC phase by the absence of a strong
anharmonic modulation, see below.

%**********************************************************************************************
%**************************************** NEW SECTION *****************************************
%**********************************************************************************************
\section{\label{sec:mic}Microscopic measurements}
\subsection{\label{sec:mic-temp}Zero-field temperature dependence}
%temperature dependence of propagation vector, comparison of heating and cooling

\begin{figure}[!t]
    \centering
    \includegraphics[width=0.45\textwidth]{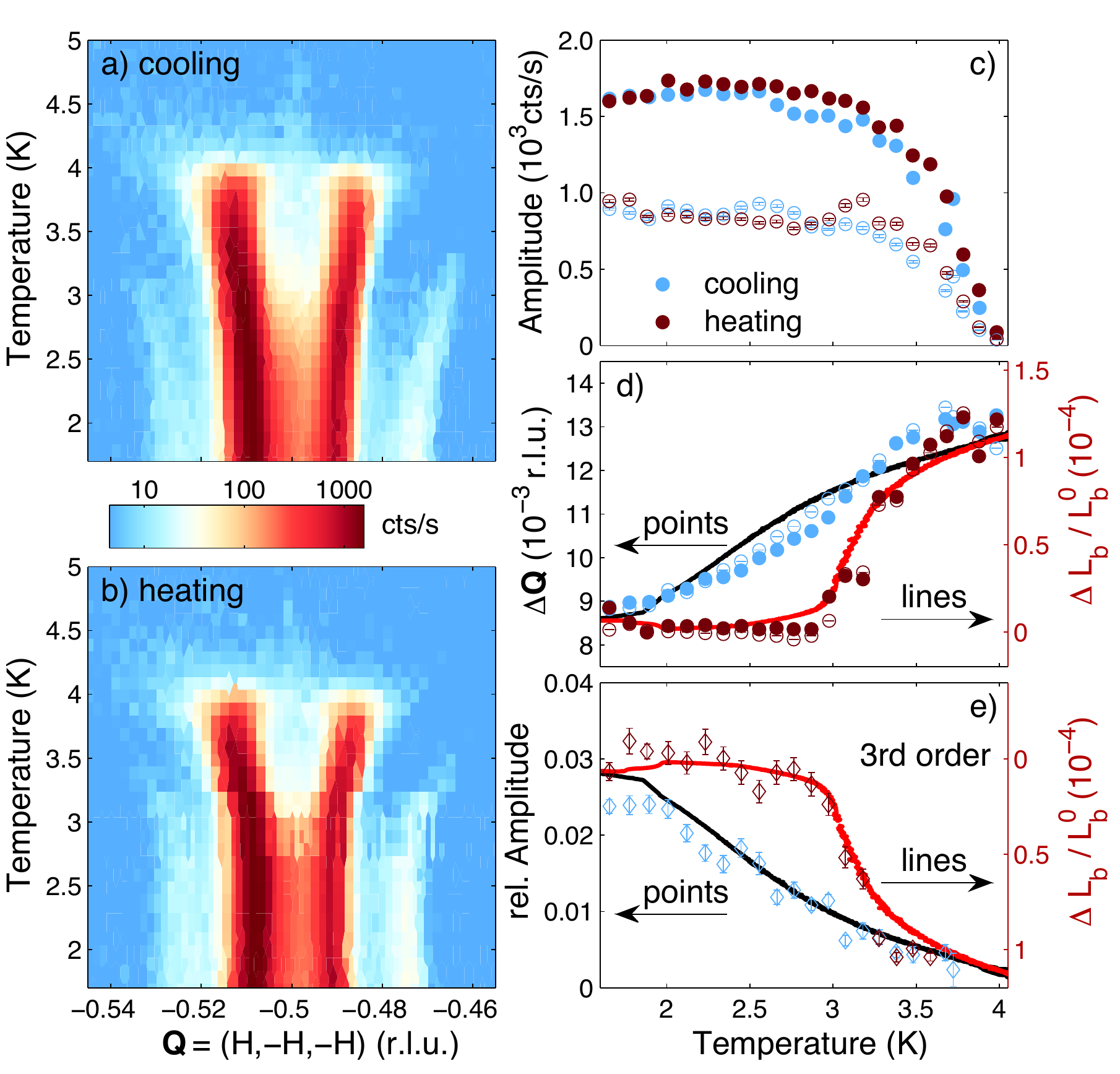}
    \caption{Magnetic phase transition in \nafewo\ upon cooling and heating at IN3. (a), (b) Intensity mapping of $\vec{Q}$ scans along the position of the magnetic Bragg peak $\vec{Q}=(-0.5,~0.5,~0.5)$ upon cooling and heating through the phase transition. The color is logarithmically coded. Gaussian peaks were fitted to the data. (c) Peak intensities of both magnetic satellites (open and closed markers, respectively). (d) positions of the satellites relative to the commensurate Bragg point, (e) Relative peak intensity of the incommensurate third-order signal upon heating and cooling. The onset of the third-order reflections coincides with the modulation of the propagation vector. The temperature dependence in (d) and (e) are compared to the thermal expansion data in zero-field (black and red lines from Fig.~3(c)).}
    \label{fig:03-heating}
\end{figure}

The temperature dependence of the magnetic superstructure
reflections was investigated  by neutron diffraction at IN3. We
worked with a fixed neutron energy of 14.7\,meV
($k_f=2.66$\,\AA$^{-1}$) and a sample orientation of
$[1,0,0]/[0,1,1]$. Figures~\ref{fig:03-heating}(a) and (b) show
intensity maps of \qvec\ scans along $[1,1,1]$ over the magnetic
satellites at $\vec{Q}=(-0.5,~0.5,~0.5)$ upon cooling and heating.
At about 4\,K, two strong incommensurate Bragg peaks develop whose
positions are temperature dependent. The magnetic satellites are
accompanied by weak third-order reflections and a weak signal
at the commensurate Bragg peak position. Well above the transition
temperature, strong diffuse scattering persists in agreement with the low-dimensional character deduced from the
macroscopic measurements. Three Gaussian
functions were fitted to the data to take into account the
intensities at the two incommensurate positions and the
commensurate position in the center. The signal at the
commensurate Bragg position is weak in comparison to the
incommensurate peaks but cannot be attributed to $\lambda/2$
contamination from a structural peak since its intensity varies
with temperature. A detailed analysis of this finding is
limited by the weakness of the signals and because the
magnetic satellites are so close to each other. This is an essential difference to \mnwo\ where the incommensurate magnetic modulation is much further away from the commensurate value. Anharmonic components were also observed in the AF2 phase of MnWO$_4$ \cite{fingerb} and there they  are related with magnetoelectric memory effects observed for the electric-field
control of multiferroic domains \cite{taniguchi,fingera}.

%Our data suggests that the commensurate signal grows towards lower temperatures which comes along with a broadening of the width of the incommensurate satellites. A development of a second magnetic phase would lower the correlation length of the primary structure which would be visible in a broadening of the magnetic reflections. The presence of a commensurate magnetic signal and higher order reflections of the incommensurate satellites are both indicators that the present magnetic structure is perturbated. \\

The appearance of the third-order reflections must be attributed
to an anharmonic perturbation of the incommensurate structure. A
deformed sinusoidal wave can be described by additional wave
vectors in the Fourier transformation. Higher-order harmonics
often indicate a squaring-up of the magnetic structure. This is an
expected feature at low temperatures as an incommensurate
sinusoidal spin-density wave cannot be the ground state of a local
moment system~\cite{Rossat-Mignod1979}. The ratio of third- and
first-order satellites is $I_{3rd}/I_{1st}\approx3$\% in \nafewo .
Figure~\ref{fig:03-heating}(c) shows the temperature dependence of
the fitted peak intensity of the first-order satellites. The rapid growth in intensity below about 3.8\,K signals the development of long-range magnetic order, in good agreement with the N\'eel temperature $T\simeq 3.9$\,K derived from the susceptibility and specific heat measurements.

%Fitting a
%power law to the data, we get the same transition temperature of
%about 3.8\,K for the cooling and heating cycle. This finding
%perfectly agrees with the temperature determined from magnetic
%susceptibility and specific heat within the temperature accuracy
%of the used cryostats. Note, however, that the zero-field thermal
%expansion data do not exhibit a strong anomaly at this temperature
%but only at lower temperatures.

In Figure~\ref{fig:03-heating}(d), the fitted peak positions of
the first-order magnetic satellites are shown for the cooling and
heating cycles. Note that here the distance $\Delta Q$ to the commensurate position is
plotted. Both first-order satellites show the same behaviour within a cycle, but the temperature dependence is different for heating and cooling. Upon cooling, the
incommensurability $\Delta Q$ decreases continuously with decreasing temperature, while upon
heating, $\Delta Q$ remains constant up to 3.0\,K and increases rapidly at higher temperatures.
An analogous temperature hysteresis is seen in the temperature dependence of the third-order magnetic satellites, see Fig.~\ref{fig:03-heating}(e) where the relative intensities
$I_{3rd}/I_{1st}$ are plotted.

The temperature hysteresis of $\Delta Q$ and $I_{3rd}/I_{1st}$ remarkably resembles the temperature hysteresis observed in the zero-field thermal expansion data $\Delta L_b(T)$, which is included in Figures 5(d) and 5(e). In contrast, the onset of incommensurate harmonic order at $T_N$ has no magnetoelastic impact on the $b$ lattice parameter. In most magnetoelastic materials\cite{magel1,magel2,magel3,magel4,magel5}, anomalies in the strain are coupled to the order parameter, typically the (staggered) magnetic moment, and thus appear just at $T_N$, while the situation in \nafewo\ is more complex. In \nafewo, the change
in the lattice is not proportional to a power of the averaged ordered moment,  $\langle \vert m \vert \rangle$, but to the emergence of anharmonicity either in the
incommensurate phase or in the commensurate order. A semi-quantitative analysis of the magnetoelastic coupling will be given in section VI after the discussion of the
commensurate magnetic structures.

\subsection{\label{sec:mic-prop}Propagation vector}
\begin{figure*}[!t]
    \centering
    \begin{minipage}[b]{0.18\textwidth}
        \centering
        \includegraphics[width=0.98\textwidth]{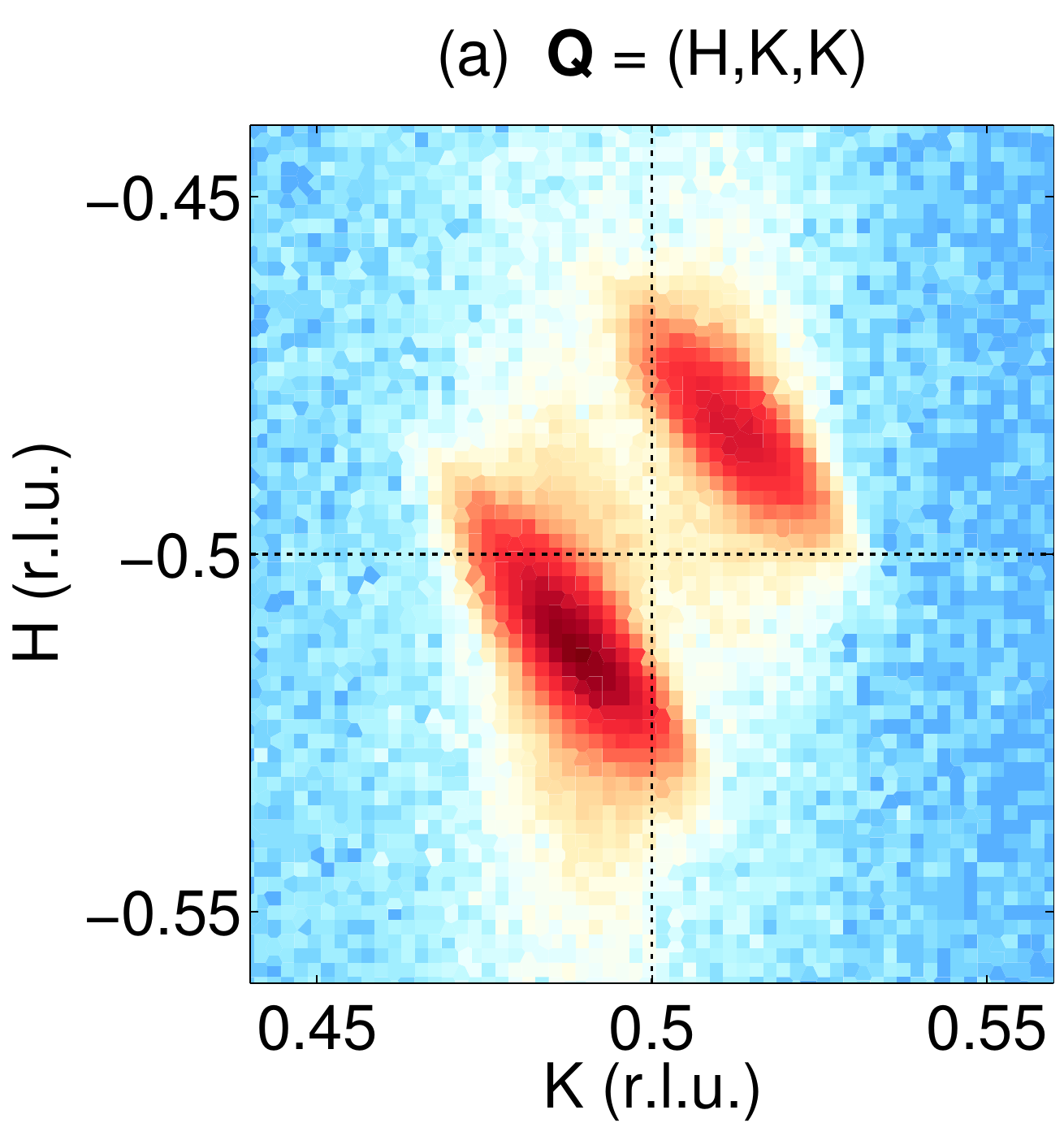}
    \end{minipage}
    %\hspace{1mm}
    \begin{minipage}[b]{0.18\textwidth}
        \centering
        \includegraphics[width=0.98\textwidth]{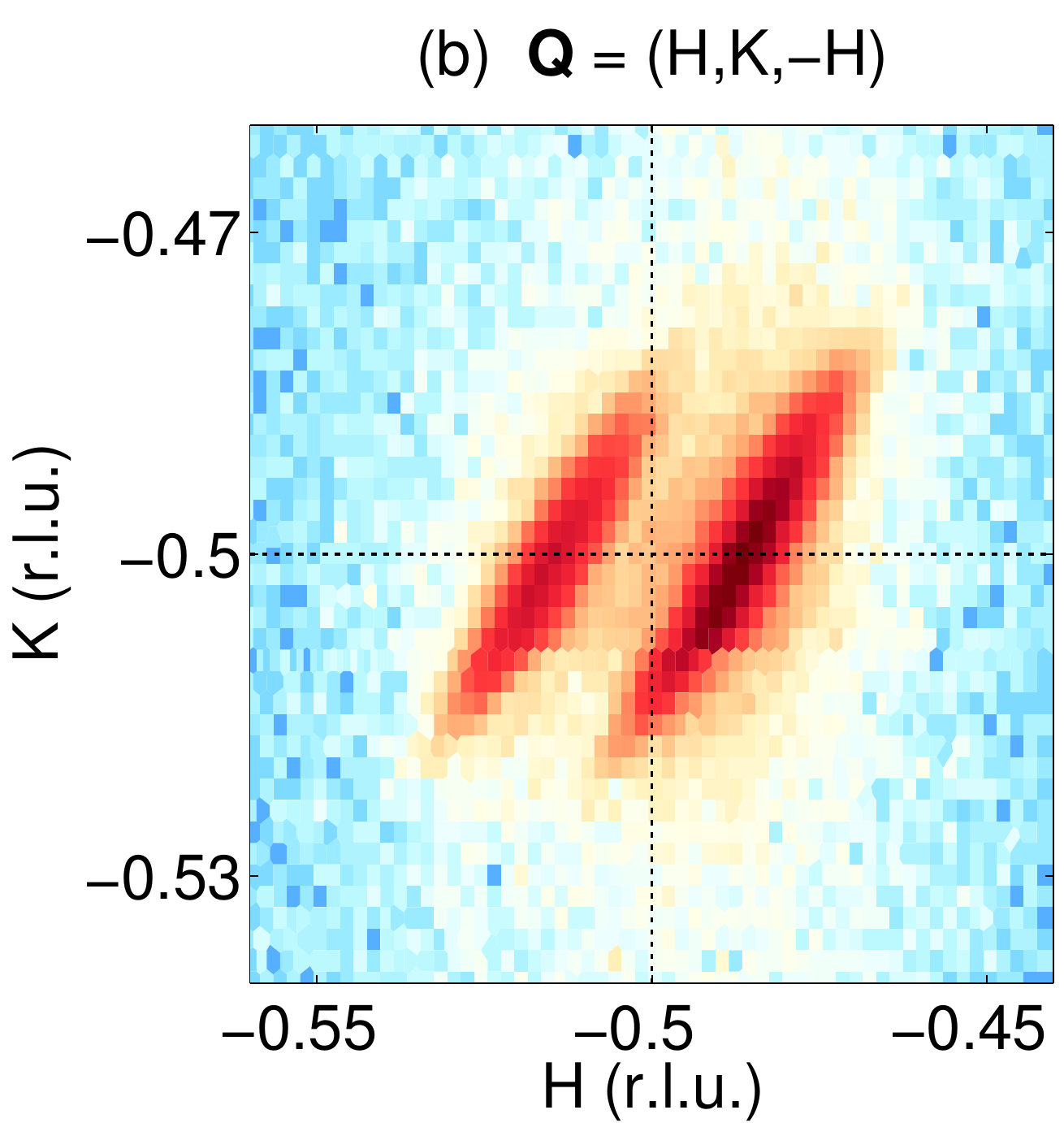}
    \end{minipage}
    \begin{minipage}[b]{0.18\textwidth}
        \centering
        \includegraphics[width=0.98\textwidth]{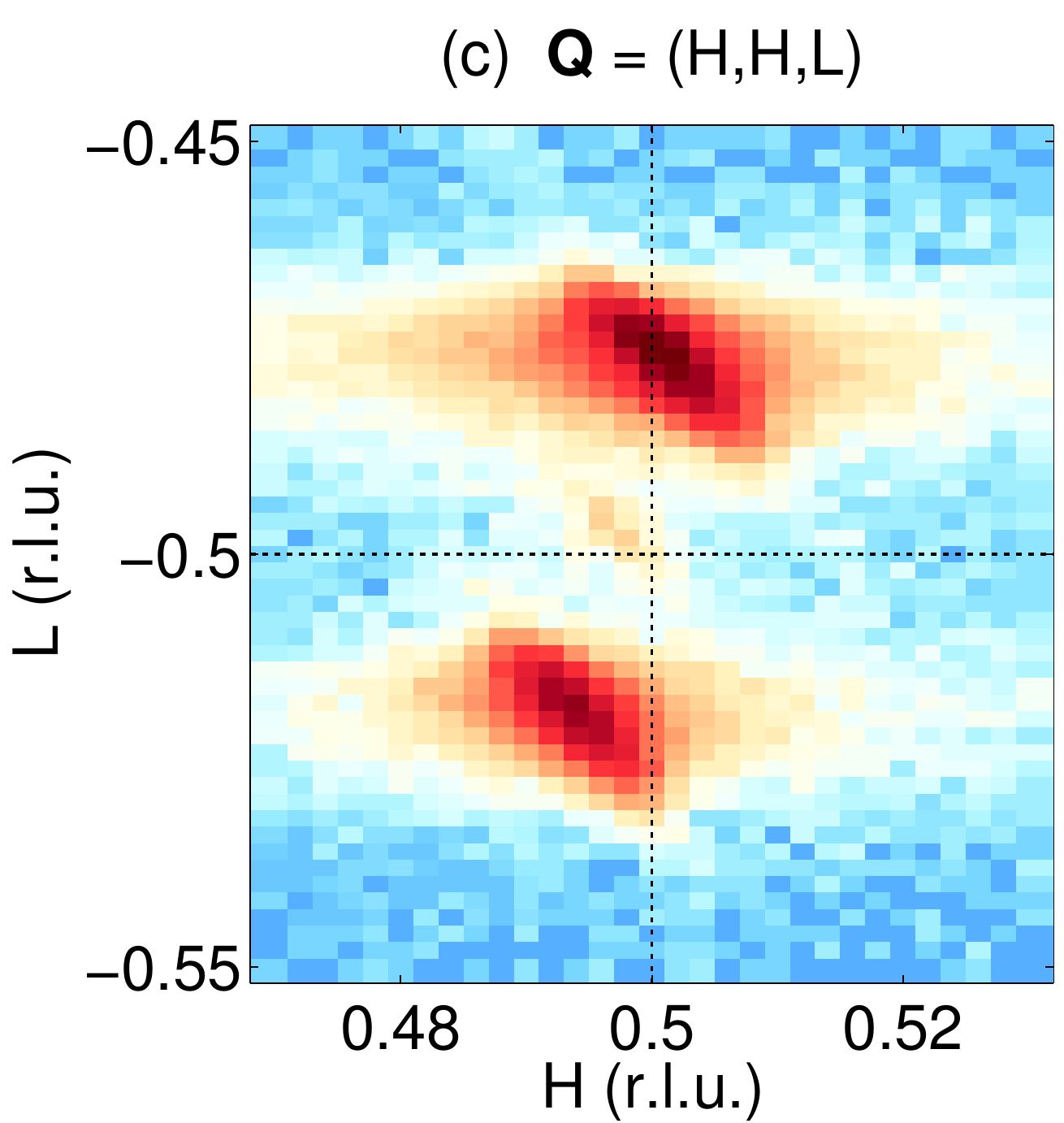}
    \end{minipage}
    \hfill
    \begin{minipage}[b]{0.42\textwidth}
    \caption{Intensity mapping of $\vec{Q}$ scans along magnetic Bragg peaks of \nafewo\ at 3.8\,K in three different orientations: (a) $\vec{Q}=(-0.5,~0.5,~0.5)$ in $[1,0,0]$/$[0,1,1]$, (b) $\vec{Q}=(-0.5,-0.5,~0.5)$ in $[0,1,0]$/$[1,0,1]$ and (c) $\vec{Q}=( 0.5,~0.5,-0.5)$ in $[0,0,1]$/$[1,1,0]$. The color is logarithmically coded. The intersection of dashed lines indicates the commensurate Bragg peak position.}
    \label{fig:nafewo_prop}
    \end{minipage}
\end{figure*}
In order to investigate the precise value of the incommensurate magnetic propagation vector along the principal crystallographic directions, different crystal orientations had been used in neutron diffraction. At IN14, we worked with a neutron energy of 3.5\,meV ($k=1.3$\,\AA$^{-1}$) and a sample orientation of $[1,1,0]/[0,0,1]$. At IN3, we worked with an neutron energy of 14.7\,meV ($k=2.66$\,\AA$^{-1}$) and a sample orientation of $[1,0,1]/[0,1,0]$ and $[1,0,0]/[0,1,1]$. Figures~\ref{fig:nafewo_prop}(a-c) show two-dimensional intensity maps of $\vec{Q}$ scans along two magnetic satellites in \nafewo\ in three different crystal orientations at 3.8\,K. The intersection of dashed lines indicates the commensurate peak position. This temperature is slightly below the magnetic transition temperature, where the splitting of the satellites is most pronounced. The intensity is logarithmically coded and diffuse scattering is visible around the static Bragg peaks. The images are two-dimensional cuts through the three-dimensional \qvec\ space. In Figure~\ref{fig:nafewo_prop}(a), the single crystal was oriented along $[1,0,0]$/$[0,1,1]$ and was measured at the spectrometer IN3. The splitting of the incommensurate magnetic satellites occurs along both axes of the scattering plane. The splitting along $[1,0,0]$ is $2\Delta_H\approx0.03$. Figure~\ref{fig:nafewo_prop}(b) was recorded at IN3 in the scattering plane $[0,1,0]$/$[1,0,1]$. In this orientation the splitting is only present along $[1,0,1]$. Within the experimental precision we cannot determine a splitting along the monoclinic axis, $\Delta_K=0$. Finally at the spectrometer IN14 the crystal was oriented along $[0,0,1]$/$[1,1,0]$. The splitting is again present along both axes and we obtain $2\Delta_L\approx0.04$ for the incommensurability along \caxis*.

We can conclude that the incommensurate splitting of the magnetic propagation vector in the zero-field phase of \nafewo\ only occurs perpendicular to the monoclinic axis \baxis. This is a symmetry plane of the Brillouin zone for the space group \ptwoc. The propagation vector at 3.8\,K is \kic.

\subsection{\label{sec:mic-diff}Diffuse scattering}
\begin{figure}[!t]
    \centering
    \includegraphics[width=0.45\textwidth]{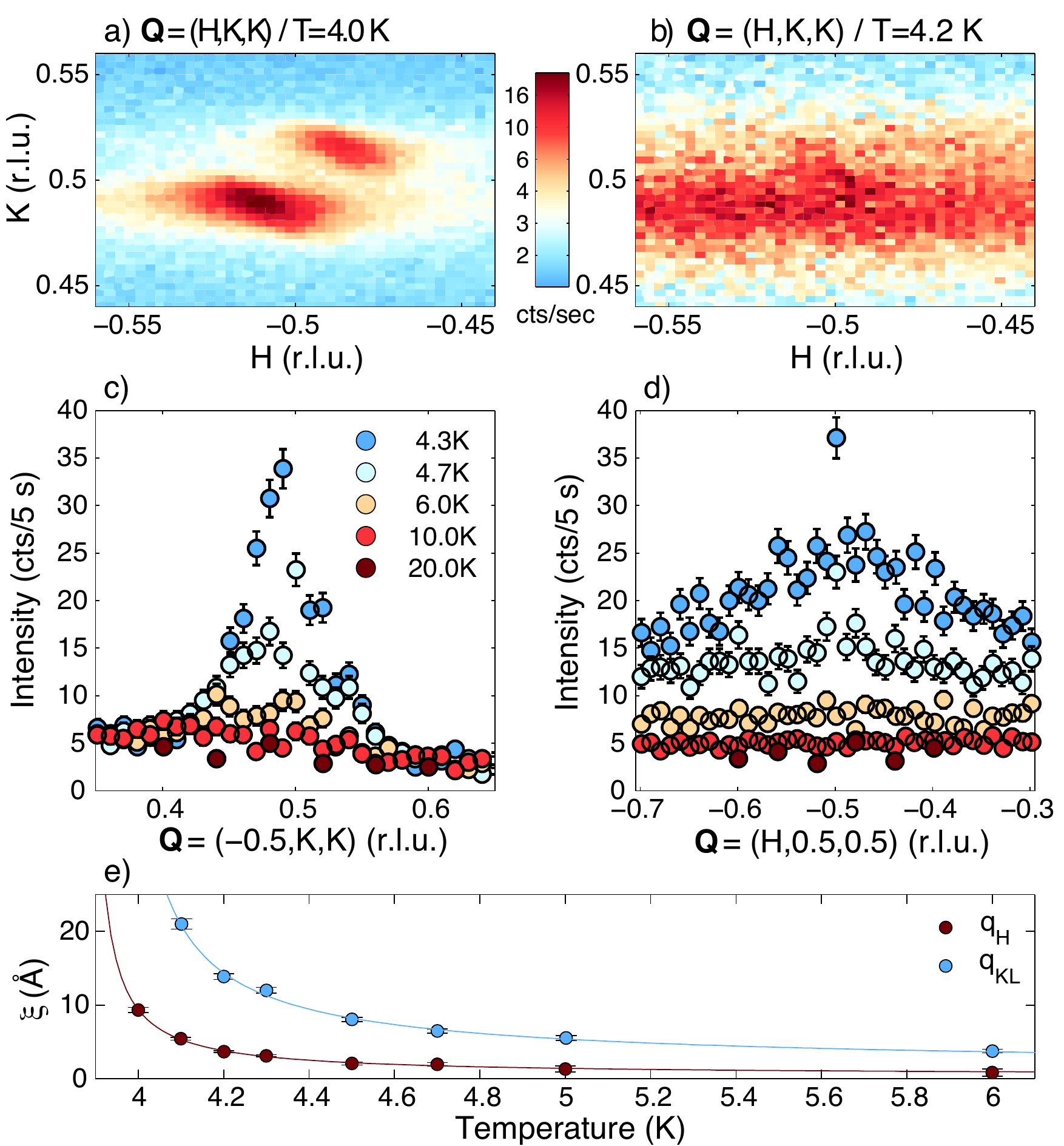}
    \vspace{5mm}
    \caption{Diffuse scattering in \nafewo. Two-dimensional intensity mapping of $\vec{Q}$ scans around $\vec{Q}=(-0.5,~0.5,~0.5)$ upon heating through the phase transition in the orientation $[1,0,0]/[0,1,1]$ at (a) 4.0\,K and (b) 4.2\,K, slightly above $T_N=3.9$\,K at IN3. The color is logarithmically coded. (c) and (d) show \qvec\ scans through the commensurate Bragg position along $[0,1,1]$ and $[1,0,0]$ at several temperatures. Diffuse magnetic scattering is present up to 6\,K. (e) Correlation length $\xi$ determined by Lorentzian fits for both directions. A power law function $\xi\propto(T-T_N)^{\nu}$ was fitted to the data.}
    \label{fig:nafewo_diff}
\end{figure}

We will continue with the investigation of the temperature and \qvec\ dependence of the diffuse scattering in the paramagnetic phase of \nafewo. The experiment was performed at the spectrometer IN3 using a single crystal of \nafewo\ in the orientation $[1,0,0]/[0,1,1]$. Figures~\ref{fig:nafewo_diff}(a) and (b) show two-dimensional intensity maps along the commensurate Bragg peak $\vec{Q}=(-0.5,~0.5,~0.5)$ at 4.0\,K and 4.2\,K, slightly above the ordering temperature. At 4.0\,K, the diffuse scattering is well centered around the incommensurate Bragg peak positions. The signal is rather sharp along $[0,1,1]$ and significantly broadened along $[1,0,0]$. At 4.2\,K this situation is even more pronounced. The diffuse scattering remains centered at the commensurate value along $[0,1,1]$ and is nearly constant along the $a$* axis.

Figure~\ref{fig:nafewo_diff}(c) shows \qvec\ scans along $[0,1,1]$ over the commensurate Bragg peak position $\vec{Q}=(-0.5,~0.5,~0.5)$ at different temperatures above the magnetic transition. Diffuse scattering is present up to 6\,K, which is 1.5 times larger than the transition temperature of $T_N=3.9$\,K suggesting a low-dimensional or frustrated character of the system. This finding is in perfect agreement with the macroscopic measurements presented in Section~\ref{sec:mac}. Magnetic resonance studies on \nafewo\ also revealed a two-dimensional character of the magnetic order and the ratio of intralayer $J$ to interlayer exchange $J'$ was estimated to be $J'\approx10^{-6}J$~\cite{Dergachev2005}.

We now focus on the anisotropy of the magnetic correlations. Figure~\ref{fig:nafewo_diff}(d) shows the temperature dependence of \qvec\ scans along $[1,0,0]$. The diffuse scattering is significantly broadened along this direction. By fitting the data with a Lorentzian function, one can determine the correlation length of the diffuse order. The finite instrument resolution can be neglected in the investigated temperature range because the diffuse signal is significantly broadened.

The temperature dependence of the correlation length along both directions is shown in Figure~\ref{fig:nafewo_diff}(e). \nafewo\ crystallizes in a layered structure with separated planes of Na, Fe and W parallel to the $bc$ plane (see Fig.~\ref{fig:nafewo_nuc} and Ref.~\cite{Klevtsov1970}). The distance of the magnetic ions along \aaxis* is almost 10\,\AA. The resulting weakness of the coupling along \aaxis* is visible in the two-dimensional diffuse scattering in the paramagnetic phase. Correlations between the magnetic moments first occur below 10\,K inside the $bc$ planes, where the magnetic moments form closely neighboring zig-zag chains. Only at lower temperatures the system develops 3-dimensional correlations between the planes. \\

\subsection{\label{sec:mic-zero}Magnetic and nuclear structure at zero field}
%crystal structure
The crystal and magnetic structure of \nafewo\ was investigated at D10. Two single crystals ($13\times8\times2$\,mm$^3$ and $6\times7\times2$\,mm$^3$) were used for the experiment. The D10 diffractometer was equipped with a $80\times80$\,mm$^2$ microstrip area detector and two wavelengths, 1.26\,\AA\ and 2.36\,\AA, were used. Magnetic Bragg reflections were recorded at 1.75\,K and structural Bragg reflections were recorded at 12\,K, well above the magnetic phase transition. The magnetic propagation vector \kic\ of \nafewo\ is incommensurate. However, the
resolution of D10 is insufficient to separate
the two satellites corresponding to $\vec{k}_{\text{ic,1}}=(0.485,~0.5,~0.48)$ and $\vec{k}_{\text{ic,2}}=(-0.485,~0.5,-0.48)$. The collection of magnetic peaks has been done by long scans at the positions in $Q$-space generated by the commensurate propagation vector. The structure refinement was done in the space group \ptwoc\ with the lattice parameters given in the introduction using the program \fullprof~\cite{progFullprof}. The datasets from both single crystals yield quantitatively the same results and we will present only the results from the more complete dataset.

Structural reflections were recorded at 12\,K in the paramagnetic phase. At a neutron wavelength of 1.26\,\AA, a total of 766 reflections were collected. For the refinement 367 independent nuclear reflections were used. The internal and weighted $R$-values are 1.64 and 1.83\%, respectively (on the intensity). The results of the refinement of the structural parameters are given in Table~\ref{tab:nafewo_FPstruc}. Isotropic temperature factors and anisotropic extinction correction (model 4 in \fullprof~\cite{progFullprof}) were applied. The values for the atomic positions correspond very nicely to the results obtained from powder data~\cite{Nyam-ochir2008} and the anisotropic extinction parameters reflect the plate-like shape of the crystal.

%One can estimate the necessity of an absorption correction for cylindrical shaped crystals by calculating the product of the linear attenuation factor of the scattered sample with the radius of the sample. In the case of \nafewo\ the attenuation factor is $\mu = 0.194$\,cm$^{-1}$ and the average diameter of the sample is approximately 0.5\,cm. For products well below a value of $1$, a correction for absorption is not necessary, which is the case for our sample~\cite{Schmitt1998}.\\
The structural dataset can also be used to verify the occupation of the different atomic sites. It was mentioned before that the mechanism of the magnetic coupling along the extended $a$ axis is still unclear. The coherent neutron scattering length of sodium and iron are $b_{\text{Na}}=3.63$\,fm and $b_{\text{Fe}}=9.45$\,fm, respectively, which renders a differentiation of both elements possible. The refinement with \fullprof\cite{progFullprof} yields a deviation of only 1 to 2\% per site. The layered structure is thus well ordered and an influence of mixed occupation on the magnetic structure can be a-priori excluded. \\
\begin{table}
    \renewcommand{\arraystretch}{1.3}
    \centering
    \caption{Structural parameters of \nafewo\ at 12\,K in the space group \ptwoc, with $a=9.88$\,\AA, $b=5.72$\,\AA, $c=4.94$\,\AA\ and $\beta=90.33$. The data was recorded at the diffractometer D10 and the refinement was done using \fullprof~\cite{progFullprof}.}
    \begin{tabular}[b]{l@{\hskip 5mm}||l|l|l@{\hskip 5mm}|l}
          & x & y & z & $U_{iso}$ (\AA$^2$) \\
        \hline
        Fe  &   0.0         & 0.67074(19)   & 0.25      & 0.04(2)   \\
        Na  &   0.5         & 0.6971(6)     & 0.25      & 0.35(5)   \\
        W   &   0.23704(14) & 0.1831(2)     & 0.2572(3) & 0.12(3)   \\
        O1  &   0.35385(12) & 0.3813(3)     & 0.3816(3) & 0.25(3)   \\
        O2  &   0.10888(13) & 0.6226(3)     & 0.5923(3) & 0.22(2)   \\
        O3  &   0.33177(13) & 0.0897(2)     & 0.9533(3) & 0.22(3)   \\
        O4  &   0.12606(13) & 0.1215(3)     & 0.5757(3) & 0.17(2)   \\
        \hline
        \multicolumn{5}{c}{$R_{F^2}=3.70$,  $R_{wF^2}=3.55$,  $R_{F}=2.87$,  $\chi^2(I)=4.13$}\\
    \end{tabular}
        \label{tab:nafewo_FPstruc}
\end{table}
%magnetic
Magnetic Bragg reflections were recorded at 2\,K in the ordered
phase. At a neutron wavelength of 1.26\,\AA, a total of 423
reflections were collected. A total of 411 independent magnetic
reflections were used for the refinement. As mentioned above, the
incommensurate satellites could not be measured independently at D10. Instead, we used
the commensurate propagation vector to measure the magnetic
reflections and integrated over both incommensurate peaks. The refinement program
\fullprof~\cite{progFullprof} allows to treat the list of measured
intensities in a way that the contribution of two neighboring
magnetic satellites is summed up in clusters and the
incommensurate propagation vector could be used for the
refinement.

\begin{table*}[!t]
    \renewcommand{\arraystretch}{1.3}
    \centering
 \caption{Residual values for the refinements of the magnetic reflections of \nafewo\ taken at 2\,K on D10 using different models. The commensurate (COM), spin-density-wave (SDW) and spiral models are explained in the text. For the spiral models, the major axes are along \easyac\ and \baxis. The best results are achieved assuming a spin spiral described by $\Gamma_2$. }
    \begin{tabular}[b]{l||r|r|r|r|r|r|r}
                    &   COM $a$   &  SDW $a$    &   SDW $ac$    & SDW $abc$     & spiral $\Gamma_1$ &   \underline{spiral $\Gamma_2$} & spiral $\Gamma_1+\Gamma_2$\\
        \hline
$R_{F^2}$   &    47.8   &   47.9    &   15.0    &   14.9    &   12.2 & 11.7 &  15.1 \\
$R_{wF^2}$  &    51.8   &   51.8    &   16.0    &   15.6    &   13.9 & 13.3 &  15.5 \\
$R_{F}$     &    29.2   &   29.2    &    9.1    &    9.1    &    7.3 &  7.1 &  9.0  \\
$\chi^2(I)$ &   239.0   &  240.0    &   22.7    &   22.1    &   10.9 &  9.9 &  21.6 \\
    \end{tabular}

\end{table*}

Different models were used to describe the data. The two sites
were described by identical Fourier coefficients and the phase
shift $\phi_{\vec{k}_{\text{c}}}$ arising from the different $z$ values. A
comparison of the refinements using different models is given in
Table~IV. The previous analysis of neutron powder data
yielded a model with a commensurate propagation vector and moments
aligned antiparallel along \aaxis~\cite{Nyam-ochir2008}. This
model, however, is not compatible with the single-crystal data
from the D10 diffractometer. The fit is improved by allowing
the spins to rotate in the $ac$ plane. Another minor improvement
can be achieved when we allow an additional component along the
monoclinic axis \baxis, that, however, remains small. This result agrees with the analysis of the
magnetic susceptibility, which suggests a magnetic moment primarily ordered
in the $ac$ plane.

The fit results are similar for a collinear spin density wave
(SDW) and a spin-spiral rotating in the \easyac-\baxis\ plane. The
vector \easyac\ denotes the direction of the easy axis in the
$ac$ plane and has an angle of $\approx48$\degree with the $a$ axis. Spin spirals with a different rotation axis are not
compatible with the data.

We now take into account the full symmetry analysis for the case
of an incommensurate propagation vector $\vec{k}_{\text{ic,1}}=(0.485,~0.5,~0.48)$, cf. Table~\ref{tab:nafewoCharIC}. Both Fe sites are
connected by a glide plane $c$ along the monoclinic axis and have
a phase difference $\phi_{\vec{k}}=2\pi\times0.24$. The
refinement gives the best result for an elliptical
spin spiral with moments rotating in the \easyac-\baxis\ plane.
This model is compatible with either of the two irreducible
representations $\Gamma_1$ and $\Gamma_2$ and refining both models
yields only slightly better reliability values for $\Gamma_2$, but the summation of neighboring magnetic satellites prohibits a clear
differentiation. This $\Gamma_2$ model is displayed in Fig.~8 and it corresponds to the
non-multiferroic AF3 phase in MnWO$_4$~\cite{Urcelay-Olabarria2013}. A
combination of both representations constrained to identical chiral
structures at both sides (only three independent parameters)
clearly worsens the fit.

%discussion of multiferroicity
The difference of the spiral models,
using either one or both irreducible representations, is the
rotation of the two moments in the crystallographic unit cell
relative to each other. When only one representation is applied,
the moments rotate in the opposite sense, relative to each other.
Only the combination of both representations allows the spirals to
rotate along the same direction, which however is necessary in
order to imply a finite ferroelectric polarization inverse \DM\ effect $\vec{D}_{ij}\times(\vec{S}_i\times\vec{S}_j)$~\cite{book_khomskii}. Such an
arrangement is for example observed in the multiferroic AF2 phase in MnWO$_4$
but not in \nafewo. This is in agreement with the absence of a pyroelectric current in \nafewo , which was reported recently~\cite{diss_ackermann}. A similar situation was discussed for the AF3 phase in \mnwo\ applying the superspace formalism~\cite{Urcelay-Olabarria2013,Solovyev2013}. Urcelay-Olabarria \textit{et al.} describe the AF3 structure as counter-rotating spirals, which prohibit the development of a ferroelectric polarization in this phase. Furthermore, also for Co-doped Ni$_3$V$_2$O$_8$ such a compensation of spiral objects with opposite signs has been reported~\cite{Qureshi2013}.\\

\begin{figure}[!b]
    \centering
    \begin{minipage}[b]{0.3\textwidth}
        \centering
        \includegraphics[width=0.7\textwidth]{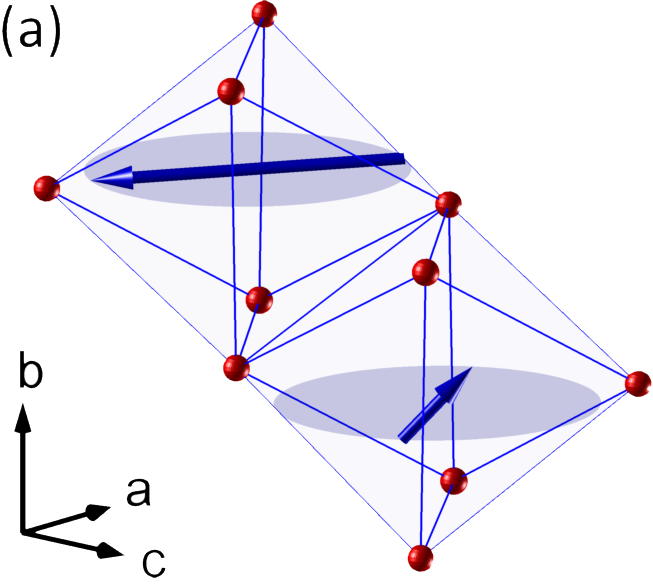}
    \end{minipage}
    \vskip5mm
    \begin{minipage}[b]{0.49\textwidth}
        \centering
        \includegraphics[width=0.99\textwidth]{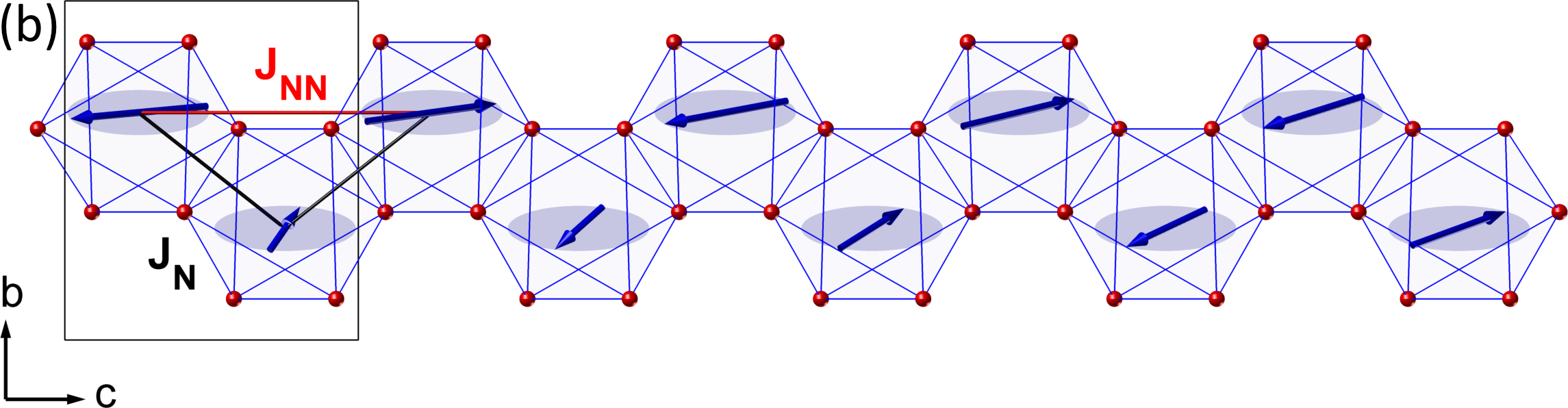}
    \end{minipage}
    \caption{Magnetic structure of \nafewo\ at 2\,K corresponding to $\Gamma_2$ as it is determined by single-crystal diffraction on D10: (a) Crystallographic unit cell with magnetic moments and oxygen ions and (b) evolution of the spiral along \caxis. The ellipses show the rotation plane of the magnetic moments with the principal axes \easyac\ and \baxis. Note, that the spirals in the upper and lower rows rotate with opposite sense.}
    \label{fig:nafewo_D10mag}
\end{figure}

%details of the model
The lengths of the major and the minor principal axis of the
elliptical spiral in Fig.~8 are $M_{max}=\sqrt{(M_x^2+M_z^2)}=4.88(4)$\mubohr\
and $M_{min}=M_y=1.09(5)$\mubohr, with a ratio
$M_{min}/M_{max}=0.22$. The angle between the major principal axis
and the $a$ axis is 47.7\degree. Given the strong
deformation of the ellipse, the magnetic moments cannot order at
every position in the lattice, similar to the case of a
spin-density wave. By comparing the area of the ellipsoid with
a circle of the same area, one obtains an average oriented moment of
about $3.5$\mubohr , which is only 70\,\% of the moment of
Fe$^{3+}$. Note, however, that the model described by \fullprof\cite{progFullprof}
only accounts for the harmonic incommensurate spin spiral. The
anharmonic squaring up which increases the ordered moment is are not taken into account in this model.

\subsection{\label{sec:mic-field}High-field phase}
\begin{figure}[!t]
    \centering
    \includegraphics[width=0.45\textwidth]{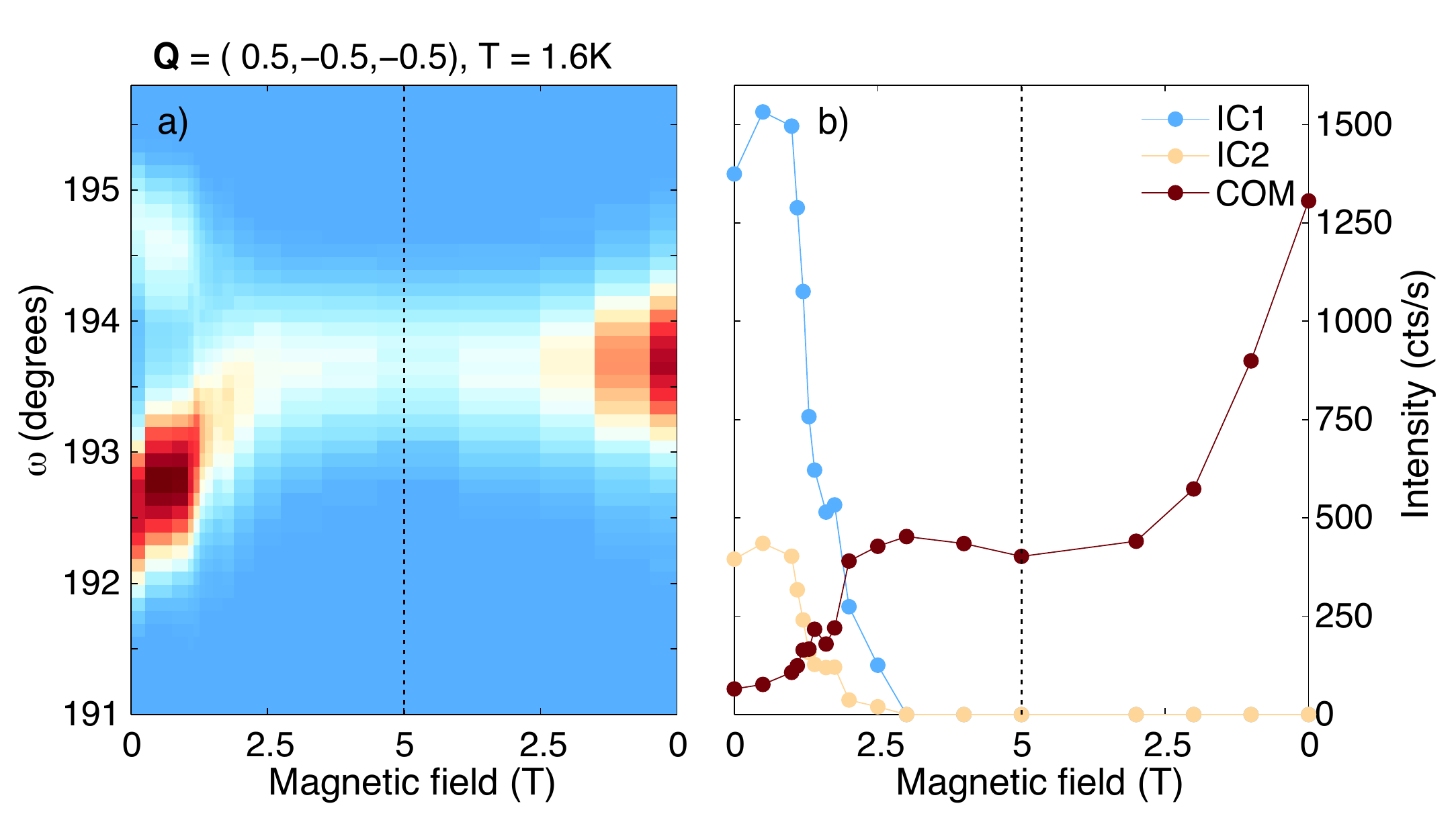}
    \caption{Magnetic field dependence of the magnetic structure in \nafewo\ at 6T2. (a) Intensity mapping of rocking scans along the magnetic Bragg peak position $\vec{Q}=( 0.5,-0.5,-0.5)$ for increasing and decreasing magnetic field $B\parallel\vec{b}$ at 1.6\,K. (b) Corresponding Bragg peak intensities of incommensurate (IC1 and IC2) and commensurate (COM) reflections fitted by Gaussian functions. Lines between points are a guide to the eye. }
    \label{fig:04-field}
\end{figure}

Finally, we investigate the high-field magnetic phase. The experiment
was performed at 6T2 with a neutron wavelength of 2.35\,\AA.
Figure~\ref{fig:04-field}(a) shows the intensity map of rocking
scans along the magnetic Bragg peak position
$\vec{Q}=(0.5,-0.5,-0.5)$ for magnetic fields applied along the
monoclinic axis \baxis. The intensity is logarithmically color
coded. The application of the magnetic field along the monoclinic
axis strongly affects the incommensurate splitting of the
satellites. At a magnetic field of about 1.2\,T the satellites
merge into one commensurate peak. The field was first increased to
a maximum field of 5\,T and then decreased to zero field at
constant temperature. In decreasing fields, only a modulation of
the intensity is visible and the scattered intensity remains at
the commensurate position. Gaussian functions have been fitted to
the data and resulting amplitudes are shown in
Figure~\ref{fig:04-field}(b). The first transition at 1.2\,T
perfectly matches the phase transition observed in the
magnetostriction data (cf. Figure~\ref{fig:02-mac}(b)).

We can thus assign three different magnetic phases in \nafewo \ as it is shown in the phase diagram in Fig. 4:
The system undergoes a phase transition from a paramagnetic towards a
low-field incommensurate (LF-IC) magnetic structure with a propagation vector
\kic. In magnetic fields applied along \baxis, the propagation vector becomes commensurate \kc\ and this phase is denoted as high-field commensurate (HF-C). Finally, the commensurate structure changes when the magnetic field decreases again which defines the low-field commensurate (LF-C) phase, see Fig.~\ref{fig:03-phase}.

The results from the zero-field diffraction data will help to
analyze the data collected in the high-field phase of
\nafewo. In addition to the temperature and magnetic field
dependence of the propagation vector, 36 magnetic reflections were
collected at 1.6\,K in 0\,T, 5\,T and again 0\,T. The instrument
was equipped with a cryomagnet, which cannot be used in combination
with  the Eulerian cradle. The movement of the sample was therefore
limited to a rotation within the scattering plane. The installed lifting
counter geometry allowed for the movement of the detector up to
30\degree\ perpendicular to the scattering plane in order to
increase the accessible \qvec\ space. The small number of
reflections and the absence of observed reflections along
\baxis\ limits the completeness of the data set. A precise
refinement of the magnetic structure is not possible but the data
gives significant information about the orientation of the moments
in the different magnetic phases.

%incommensurate 0T
The 6T2 zero-field data confirms the model for the magnetic structure determined from the D10 data in the LF-IC phase. As a result we get the same incommensurate spin spiral with main axes along \easyac\ and \baxis, which can be described by one irreducible representation. The ratio between the components along \caxis\ and \aaxis\ is $M_z/M_x\approx1.1$, which corresponds to an angle of 47.7\degree\ to the $a$ axis. A model of the magnetic structure is shown in Figure~\ref{fig:nafewo_D10mag}(a).

\begin{figure}[!t]
    \centering
    \begin{minipage}[b]{0.235\textwidth}
        \centering
        \includegraphics[width=0.95\textwidth]{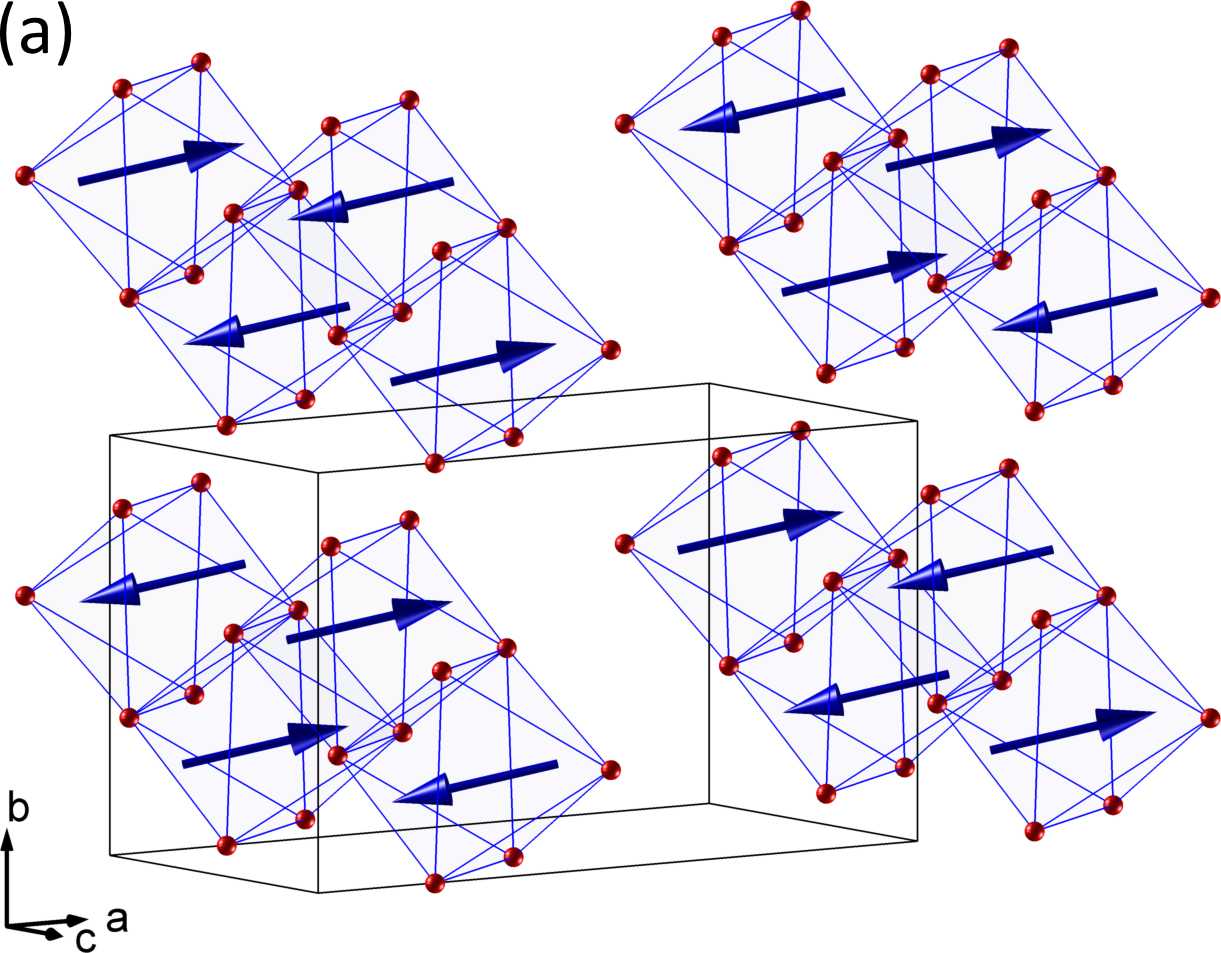}
    \end{minipage}
    \hfill
    \begin{minipage}[b]{0.235\textwidth}
        \centering
        \includegraphics[width=0.95\textwidth]{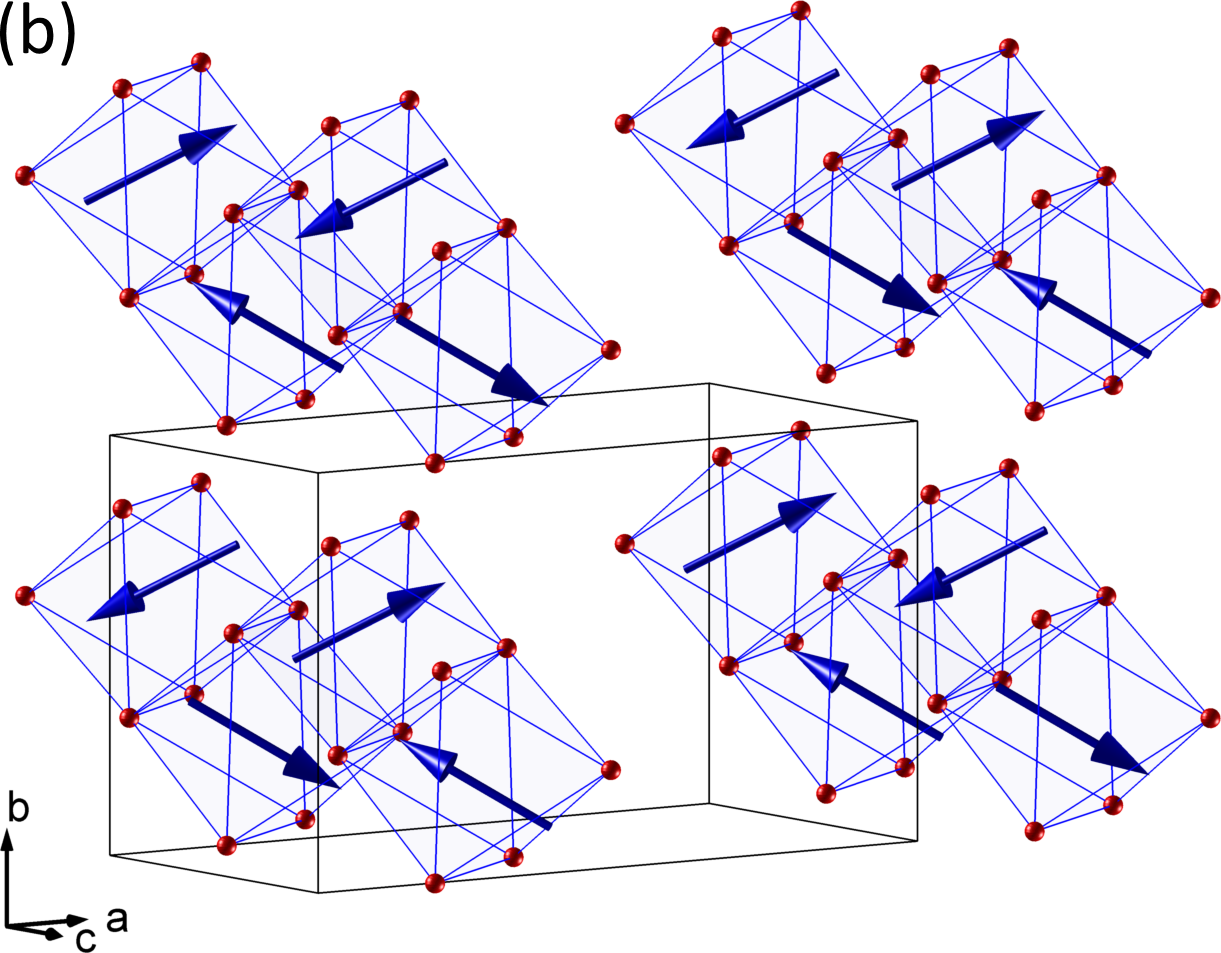}
    \end{minipage}
    \caption{Magnetic structures of \nafewo\ at 1.6\,K as determined by single crystal diffraction on 6T2 with magnetic fields applied along \baxis: (a) High-field commensurate phase at 5\,T and (b) low-field commensurate phase at zero field. The induced ferromagnetic magnetization was not taken into account.}
    \label{fig:nafewo_6T2mag}
\end{figure}

%commensurate 5T
Figure~\ref{fig:nafewo_6T2mag}(a) shows a model of the magnetic
structure determined from the 6T2 data in a magnetic field of 5\,T
applied along the monoclinic axis. The figure displays only the commensurate antiferromagnetic order without the induced
ferromagnetic moment along \baxis\ (see also
Table~\ref{tab:nafewo_FPmagall}). The propagation vector of the
HF-C phase changes to a commensurate \kc.
The relative orientation of the spin at the second Fe$^{3+}$ site in the crystallographic unit
cell was determined by a comparison of four different
configurations as described previously: $(\bar{u},\bar{v},\bar{w})$, $(\bar{u},v,\bar{w})$, $(u,\bar{v},w)$ and
$(u,v,w)$. The resulting R-values (16, 25, 52 and 96\%, respectively) show that the best fit is obtained for the antiparallel alignment. The
moments are collinear and antiferromagnetically ordered with
components along all three axes: $M_x=2.54(10)$\mubohr,
$M_y=0.84(14)$\mubohr\ and $M_z=2.12(10)$\mubohr\ yielding a total
ordered moment of $M=3.4(2)$\mubohr. This model only describes the
components ordered antiferromagnetically according to the
commensurate propagation vector. We can deduce from the
magnetization at 2\,K (cf. Fig.~\ref{fig:nafewo_SQUID}(c)) that
the field applied along \baxis\ induces an additional
ferromagnetic moment of about $M_{\text{FM}}=1.3$\mubohr\ further
enhancing the total ordered moment. This
ferromagnetic moment leads to an increase of intensity at the
nuclear Bragg peak positions. The ratio between the components
along \caxis\ and \aaxis\ amounts to $M_z/M_x\approx0.8$, which
corresponds to an angle of 39.7\degree\ to the $a$ axis. This
value is significantly smaller than in the incommensurate zero-field phase.

%commensurate 0T
Finally, Figure~\ref{fig:nafewo_6T2mag}(b) shows the model of the
magnetic structure determined with the zero-field data directly
after decreasing the field from 5 to 0\,T at a constant
temperature of 1.6\,K. The propagation vector in the LF-C phase is
the same as in the high-field phase, \kc.
The system thus does not transform back to the low-field
low-temperature LF-IC phase that is reached upon zero-field
cooling. This resembles observations in multiferroic $R$MnO$_3$,
which also exhibit a first-order phase transition from an incommensurate
to a commensurate magnetic state upon enhancing the magnetic field
and which also does not fall back to the initial magnetic
structure after full release of the field \cite{senff,Baier2006}. In
contrast to $R$MnO$_3$, there is only one element carrying a magnetic moment in
\nafewo, which documents that such hysteresis effects can just
arise from pinning due to anharmonicity and single-ion anisotropy.

The same comparison of models was performed as described before
and the best fit for the LF-C structure was achieved with canted
moments in the crystallographic unit cell with components along
all three axis: $M_x=2.5(3)$\mubohr, $M_y=2.1(2)$\mubohr\ and
$M_z=3.1(4)$\mubohr\ yielding a total ordered moment of
$M=4.5(6)$\mubohr, close to the expected value. The ratio between the components along \caxis\ and
\aaxis\ amounts to $M_z/M_x\approx1.2$, which corresponds to an
angle of 50.7\degree\ to the $a$ axis. This value is similar to
the one in the high-field phase. The transition from the HF-C to
the LF-C phase is visible in a modulation of scattered intensity
(cf. Fig.~\ref{fig:04-field}(b)) and as a spin-flop transition in
the magnetization data (cf. Fig.~\ref{fig:nafewo_SQUID}(d)). A
comparison of the models in the incommensurate low-field phase
(IC-LF), commensurate high-field phase (C-HF) and the commensurate
low-field phase (C-LF) is given in
Table~\ref{tab:nafewo_FPmagall}.

\begin{table*}[!t]
    \renewcommand{\arraystretch}{1.3}
    \centering
        \caption{Comparison of magnetic structures in \nafewo\ in magnetic fields applied parallel \baxis. The orientation of the moments in the crystallographic unit cells are given and the corresponding propagation vectors are \kc\ (C) and \kic\ (IC). The models were determined from experiments at D10 and 6T2 using \fullprof~\cite{progFullprof}. We list the Fourier coefficients $\vec{M}$ for site 1 and the symmetry relation for site 2, with $a=e^{-2\pi i\cdot k_z \cdot r_z}$. The induced ferromagnetic magnetization, $M_{\text{FM}}(B)$, is deduced from SQUID data.}
%    \begin{minipage}[b]{0.6\textwidth}
    \begin{tabular}[b]{c||c|c|c|c|c|c}
        $B || b$ (T)    &   $M_{\text{FM}}(B)$ ($\mu_B$) & phase   & site 1 $\vec{M}$ ($\mu_{\text{B}}$)    & site 2  & $\angle(\vec{e}_{ac},\vec{a})$ (\degree) & $M_{\text{tot}}$ ($\mu_B$)   \\
        \hline
        0   &   0\phantom{.00} & LF-IC  & (3.3,-i$\cdot$1.1,~3.6)       & $a\cdot(u,\bar{v},w)$         &   47.7 & 3.5\\
        \hline
        5   &   1.28           & HF-C   & (2.5,~0.8,~2.1)               & $(\bar{u},\bar{v},\bar{w})$   &   39.7& 3.6\\
        3   &   0.75           & HF-C   & (3.0,~0.8,~3.0)               & $(\bar{u},\bar{v},\bar{w})$   &   44.8& 4.4\\
        2   &   0.50           & HF-C   & (3.0,~0.5,~3.3)               & $(\bar{u},\bar{v},\bar{w})$   &   48.1& 4.5\\
        \hline
        1   &   0.25           & LF-C   & (2.7,~1.5,~3.3)               & $(u,\bar{v},w)$               &   50.4& 4.5\\
        0   &   0\phantom{.00} & LF-C   & (2.5,~2.1,~3.1)               & $(u,\bar{v},w)$               &   50.7& 4.5\\
    \end{tabular}
%    \end{minipage}
%    \hfill
%    \begin{minipage}[b]{0.38\textwidth}

    \label{tab:nafewo_FPmagall}
%    \end{minipage}
\end{table*}

%thermal expansion - anharmonic
Common to both commensurate structures, LF-C and HF-C, is the up-up-down-down arrangement of spins along the zig-zag chains parallel \caxis. Such a structure can be considered as highly anharmonic as ferro- and antiferromagnetic neighbors alternate. The sizable magnetoelastic effect at the magnetic transition, which was observed in the thermal expansion data can be directly related to this anharmonic modulation. The AF1 ground state of isostructural \mnwo\ also develops up-up-down-down chains of spins~\cite{Lautenschlager1993}. The transition towards this magnetic structure in MnWO$_4$ is also accompanied with a drastic change in the thermal expansion, but its strength is reduced by an order of magnitude relative to \nafewo \cite{Chaudhury2008}.

%**********************************************************************************************
%**************************************** NEW SECTION *****************************************
%**********************************************************************************************
\section{\label{sec:magel}Magnetoelastic coupling in NaFe(WO$_4$)$_2$}

Two aspects of the above described magnetoelastic coupling are astonishing. Firstly, the effects are rather large yielding a relative reduction of the $b$ lattice parameter by up to $\frac{\Delta b}{b}\simeq -2.6\cdot 10^{-4}$, and, secondly, there is no magnetoelastic anomaly at the onset of magnetic ordering in zero field, which results in the incommensurate phase. In most systems with strong magnetoelastic effects
one may couple the strain, $\epsilon$, with some power of the ordered moment defined as $m_{av}=\langle \vert m \vert \rangle$ \cite{magel1,magel2,magel3,magel4,magel5,feas}, but in
\nafewo\ the intermediate incommensurate phase renders the analysis more complex. Apparently there is only a weak coupling to the incommensurate phase, while that
to the commensurate order parameter is strong.

A deeper insight in the magnetoelastic coupling can be obtained from the Landau theory incorporating
powers of the order parameter and strain terms to the expansion of the free energy \cite{cowley}.
For the commensurate ordering with doubling of the lattice along $a$ and $c$ directions, which corresponds to the up-up-down-down scheme in the
chains, a linear quadratic coupling term in the free energy is allowed, because $\eta_{com}^2 $ corresponds to the zone center.
One may thus write the strain-dependent part of the free energy as \cite{cowley}:
\begin{equation}
F_{strain}=a\cdot \epsilon \eta_{com}^2 + C\epsilon ^2,
\label{lt}
\end{equation}
where $a\cdot \epsilon \eta_{com}^2$ describes the coupling between strain and the commensurate order parameter and $C\epsilon ^2$ the purely elastic energy of the deformation.
Minimizing the free energy with respect to $\epsilon$, i.e. $\frac{\partial F}{\partial \epsilon}$=0, yields the common
proportionality between the strain and the square of the order parameter \cite{cowley}:
\begin{equation}
\epsilon \propto \eta_{com}^2.
\label{sq}
\end{equation}

In \nafewo, we do not observe a second-order phase transition to the commensurate phase, but the qualitative prediction of the Landau theory
remains correct for small structural deformations also in case of first-order transitions. We may thus qualitatively understand the structural anomalies when entering the commensurate phase.

When extending this simple Landau theory one should first include the incommensurate character of the structural distortion,
which requires taking into account two order parameter components (corresponding to the incommensurate wave vectors $q$ and $-q$),
which, however, can be transformed to an amplitude, $A$,  and a phase, $\Phi$. Only the amplitude couples to the strain, in the same way as described by equation (2) \cite{cowley}.
The competition between incommensurate and
anharmonic or commensurate structural phases can be modeled by the Umklapp terms yielding a contribution $VA^p\cos(p\Phi)$ to the free energy\cite{cowley}. Here, the commensurate wave vector is $\frac{1}{p}{\bf G}$ with ${\bf G}$ a reciprocal lattice vector, and $V$ is the energy constant. In our case this will yield a quadratic term which can be expected to be strong. There are purely structural systems displaying sequences of incommensurate and commensurate phases \cite{blinc}, and for example thermal expansion measurements on Rb$_2$ZnCl$_4$ found stronger anomalies at the incommensurate to commensurate transition \cite{blinc} somewhat similar to our observation. However, in Rb$_2$ZnCl$_4$ there still is a sizeable anomaly at the incommensurate transition, and the integrated
length change in the incommensurate phase is even larger than that at the transition to commensurate order. The main shortcoming
of the Landau theory to describe  NaFe(WO$_4$)$_2$ consists in its magnetic character and the fact that the transitions between the various phases
appear when ordered moments are sizeable and close to the saturation values. Therefore a  microscopic magnetic model and its coupling to structural distortions are needed to
describe the transition between the different phases in  NaFe(WO$_4$)$_2$.
It is worth emphasizing that the sequence of magnetic transitions in NaFe(WO$_4$)$_2$ resembles that in $RE$MnO$_3$. For decreasing ionic radius of the $RE$ ion in $RE$MnO$_3$ the magnetic structure changes from an incommensurate cycloid at
$RE$=Tb or Dy to an up-up-down-down structure at smaller $RE$ \cite{mf-general}.

For the microscopic magnetic model, we restrict to the nearest-neighbor, $J_N$, and next-nearest neighbor interactions, $J_{NN}$, along the chains in a simple Heisenberg Hamiltonian
\begin{equation}
H=\sum_{\langle i,j\rangle}J_{i,j}S_i\cdot S_j,
\label{hh}
\end{equation}
see Fig. 8 (b). It is obvious that the commensurate structure satisfies an antiferromagnetic $J_{NN}$. However, the nearest-neighbor coupling $J_N$ remains fully frustrated,
so that there is no effective coupling between the upper and lower rows of the zigzag chain shown in Fig.~8(b). This resembles the $J_1$/$J_2$ frustrated square two-dimensional Heisenberg AFM model with nearest and next-nearest neighbor interaction which results in a frustration lifting distortion \cite{chandra,melzi}.
Adding intra-chain terms does not lift this frustration in \nafewo.
One may also deduce from the simple commensurate structure that the structural symmetry must become triclinic. The up-up and down-down pairs
in the zigzag chains point either along the $bc$ or the $b\bar{c}$ diagonal, see Fig.~1(b), so that the two-fold axis along $b$ is broken. The dominating
$J_{NN}$ will enforce the commensurate up-up-down-down magnetic structure, but the persisting frustration of $J_N$ is lifted by a triclinic distortion, which seems to
be coupled with the larger and therefore visible effect in the $b$ lattice parameter. This lifting of a degenerate state by a structural and ferroelastic distortion is
rather common; it has been reported e.g. for the $J_1$/$J_2$ frustrated square two-dimensional Heisenberg AFM model \cite{melzi}, for VOCl \cite{magel4}, BaMn$_2$O$_3$ \cite{magel5}, BaCo$_2$V$_2$O$_8$ \cite{bacovo} and the parent materials of FeAs based superconductors \cite{feas}.
Since the symmetry conditions are the same for the low-temperature commensurate phase of MnWO$_4$ the same analysis can be applied, and indeed magnetoelastic
anomalies were also reported for this material but they are much smaller than in \nafewo\ \cite{Chaudhury2008}. The up-up-down-down magnetic structure should also result in some
atomic displacements following either the parallel or antiparallel alignments. The determination of these displacements requires a dedicated structural analysis in the commensurate phase, which can only be reached by applying a magnetic field.
Such a structural modulation would furthermore resemble the dimerisation at the spin-Peierls transition in CuGeO$_3$, which also shows huge effects in the macroscopic strain
parameters as well as soliton effects\cite{cugeo3,Lorenz1998,Ruckamp2005,buchner}. In contrast to the up-up-down-down case, there is no frustration left in the incommensurate magnetic structure, which is reflected by the fact that there is a single
Fe orbit in this magnetic symmetry analysis, see Table II. There is thus no need for a structural distortion to lift magnetic frustration, which seems the reason
for the absence of strong magnetoelastic coupling.

\nafewo\ also exhibits a magnetoelastic anomaly when the incommensurate structure becomes anharmonic, and the change in the $b$ lattice parameter scales
well with the intensity of the third-order satellite, see Fig.~5(e). The variation of the incommensurability seems to couple with the anharmonicity and therefore also
scales with the length changes.  The length change at the transition to an anharmonic incommensurate structure can be best understood when the incommensurate phase is described within a soliton-like model with anti-phase domains and a varying order-parameter amplitude $\eta(x)$ that is either plus or minus $\eta_{com}$. The small anharmonic
modulation thus implies regions with commensurate order. Note that the incommensurability, i.e. the deviation from the commensurate propagation
vector in \nafewo , is very small, so that the modulation length or the soliton distance amounts to about 50 lattice constants. Therefore the
induced commensurate ordering results in qualitatively the same reduction of the $b$ lattice parameter. From Fig.~\ref{fig:03-heating}, one can see that the overall contraction $\frac{\Delta b}{b}\simeq -1.3\cdot 10^{-4}$  between about 4 and 0.5~K is of similar magnitude than that of the field-induced contraction $\frac{\Delta b}{b}\simeq -1.5\cdot 10^{-4}$ at the lower critical field $B_{c1}^{up}\simeq 3.8\;$T and low temperature, whereas a significantly larger contraction $\frac{\Delta b}{b}\simeq -2.6\cdot 10^{-4}$ takes place in a field of 7~T upon cooling. At high field, \nafewo\ directly transforms from the paramagnetic to the commensurate order, while this transition
is split into two steps upon cooling and subsequent ramping up the field. The sum of the length changes at the latter two transitions nicely agrees with that at cooling in high fields.

%**********************************************************************************************
%**************************************** NEW SECTION *****************************************
%**********************************************************************************************
\section{\label{sec:conc}CONCLUSIONS}

The double tungstate \nafewo\ structurally resembles the well
studied spiral multiferroic \mnwo and its magnetic structure exhibits analogies with
that in the $RE$MnO$_3$ series where incommensurate cycloid and commensurate up-up-down-down
phases compete. The magnetic phase diagram of
\nafewo\ was investigated in detail. An analysis of
the different magnetic structures by neutron diffraction together with the study of the
complex temperature and magnetic field dependence of the
propagation vector explains the magnetic phase diagram and the
strong signature of magnetic phase transitions in various
macroscopic measurements.

At zero magnetic field, the Fe$^{3+}$ magnetic moments order directly in
a spin spiral with an incommensurate propagation vector
\kic\ at 3.9\,K. The spiral is
elliptically distorted with the major axis of the spiral pointing
along \easyac\ and the minor axis along \baxis. This phase can be
described by a single one-dimensional irreducible representation.
The incommensurability decreases with temperature and freezes in at
a temperature of 2.0\,K. Upon heating, the incommensurability shows
a hysteresis behaviour, which is coupled to an anharmonic
distortion of the spiral. The hysteresis effects of the
propagation vector and of the anharmonic distortion explain strong
anomalies visible in thermal expansion data, whereas the
antiferromagnetic transition itself is almost invisible in thermal
expansion.

The direct transition into the spiral state contrasts to other
systems such as \mnwo\ and \tbmno, where the spiral phase follows
a primarily sinusoidal modulated
phase~\cite{Lautenschlager1993,Kenzelmann2005}. The spiral
transition can be described by a single irreducible
representation, which perfectly explains the absence of an
electric polarization in this phase in \nafewo. Spin spirals of
opposite rotation sense are equally present in the system and
cancel out the emergence of a macroscopic ferroelectric
polarization as described by the inverse \DM\
coupling. In contrast, for \mnwo\ the spiral state is described by a combination
of two representations, which allows for the unique chirality that
induces a finite electric polarization.

%high field
In magnetic fields applied along \baxis, the magnetic structure
becomes commensurate with a propagation vector
\kc. The collinear ordered magnetic
moment possesses components within the monoclinic plane, as well
as perpendicular to it. When the field is decreased while keeping
the temperature constant, the magnetic order shows a spin-flop
transition. The propagation vector remains commensurate but the
magnetic moments are canted. This phase is similar to the
commensurate ground state of \mnwo~\cite{Lautenschlager1993} but
clearly differs from the low-temperature incommensurate phase that is reached in
\nafewo \ upon zero-field cooling. The transition from incommensurate to commensurate magnetic order is accompanied by strong magnetoelastic anomalies, similar to those associated with the emergence of anharmonic components in the LF-IC phase. This similar magnetoelastic response can be explained  by the character of the strongly anharmonic phase, which corresponds to commensurate parts separated by a regular arrangement of domain walls.
Two aspects of the magnetoelastic coupling are remarkable: It is restrained to the commensurate, schematically up-up-down-down structure, and it is very strong yielding a relative length change of up to $\frac{\Delta b}{b}$=2.6$\cdot 10^{-4}$.

%ground state
From the magnetic phase diagram we can assume that the magnetic
ground state of \nafewo\ is the commensurate low-field phase with
an almost fully ordered moment while the LF-IC phase is metastable and exhibits an ordered moment significantly below that expected for $S=5/2$.
Upon cooling, the system first orders in the incommensurate
structure with a sizable anharmonic distortion developing below
$\sim$3\ K. But even on further cooling the system does not
transform to the LF-C phase. Applying magnetic fields along
\baxis\ at low temperature induces the transition into the
commensurate state, which persists even after full reduction of the field.

Overall the phase diagram of \nafewo \ is governed by the
interplay of anharmonic distortions and structures with the
single-ion anisotropy. For Fe$^{3+}$, the single-ion anisotropy is expected to be small but significant contributions were also observed in \textit{AB}FeO$_4$ (with \textit{A}=La,Pr and \textit{B}=Sr,Ca)~\cite{Qureshia,Qureshib}. In \nafewo, the impact of the Fe$^{3+}$ single-ion anisotropy seems enhanced by the weakness of the magnetic exchange.
Pinning of anharmonic modulations should furthermore be
relevant for the understanding of the magnetoelectric memory and
switching behavior of closely related multiferroics.

%**********************************************************************************************
%**************************************** NEW SECTION *****************************************
%**********************************************************************************************
\section*{\label{sec:ackn}ACKNOWLEDGMENTS}
This work was supported by the Deutsche Forschungsgemeinschaft
through the Bonn Cologne Graduate School and
through CRC 1238 projects A02, B01 and B04.

\end{document}